\documentclass[prb,twocolumn,aps,showpacs,superscriptaddress,floatfix]{revtex4-1}

\usepackage{amsmath}

\usepackage[dvips]{graphicx}
\usepackage{dcolumn}
\usepackage{bm}
\usepackage{latexsym}
\usepackage{amssymb}
\usepackage{times}
\usepackage[normalem]{ulem}
\usepackage[usenames,dvipsnames]{color}

\newcommand{\be}{\begin{equation}}
\newcommand{\ee}{\end{equation}}
\newcommand{\bea}{\begin{eqnarray}}
\newcommand{\eea}{\end{eqnarray}}

\def\nn{\nonumber\\}

\def\fr#1{(\ref{#1})}

\def\eps{\epsilon}

\usepackage[colorlinks,bookmarks=false,citecolor=blue,linkcolor=red,urlcolor=blue]{hyperref}

\begin{document}

\title{Phase diagram and continuous pair-unbinding transition of the bilinear-biquadratic $S=1$ Heisenberg chain in a magnetic field}
\author{Salvatore R. Manmana}
\altaffiliation[Present address: ]{JILA (University of Colorado and NIST), and Department of Physics, CU Boulder, CO 80309-0440, USA.}
\affiliation{Institut de Th\'eorie des Ph\'enom\`enes Physiques, EPF Lausanne, CH-1015 Lausanne, Switzerland}
\author{Andreas M. L\"auchli}
\affiliation{Max-Planck Institut f\"ur Physik komplexer Systeme, N\"othnitzer Stra{\ss}e 38,
  D-01187 Dresden, Germany}
\author{Fabian H.L. Essler}
\affiliation{The Rudolf Peierls Centre for Theoretical Physics, Oxford University, Oxford OX1 3NP, United Kingdom}
\author{Fr\'ed\'eric Mila}
\affiliation{Institut de Th\'eorie des Ph\'enom\`enes Physiques, EPF Lausanne, CH-1015
  Lausanne, Switzerland}
\date{\today}


\begin{abstract}
We investigate the properties of the Heisenberg $S=1$ chain with bilinear and biquadratic interactions in a
magnetic field using the Density Matrix Renormalization Group, Bethe ansatz and field theoretical considerations.
In a large region of the parameter space, we identify a magnetized ferroquadrupolar Luttinger liquid consisting
of a quasi-condensate of bound magnon pairs. This liquid undergoes a continuous pair unbinding transition to a
more conventional Luttinger liquid region obtained by polarizing the system above the Haldane gap region.
This pair unbinding transition is shown to be in the Ising universality class on top of a Luttinger liquid, leading to
an effective central charge 3/2.
We also revisit the nature of the partially polarized Luttinger liquid around and above the Uimin-Lai-Sutherland point.
Our results confirm that this is a two-component liquid and rule out the formation of a single-component vector chiral phase.
\end{abstract}

\maketitle

\section{Introduction}
\label{sec:intro}

A particularly interesting aspect of quantum many body systems is their ability to host ordered phases which emerge from the spontaneous breaking of one or several of the symmetries of the system.
In the case of spin systems, the main focus lies on possible long-range-order (LRO) due to the breaking of the SU(2) symmetry inherent to these systems.
For $S=1/2$ Heisenberg systems, breaking this continuous symmetry with a purely local order parameter
implies magnetic ordering.
However, when the spin is $S>1/2$, there is the alternative possibility that the SU(2) symmetry is broken by
a local quadrupolar order parameter, leading to spin-nematic phases.\cite{spin_nematics1,spin_nematics2}
Recent findings on NiGa$_2$S$_4$, a spin-1 material on a triangular lattice, \cite{SatoruNakatsuji09092005} indicate the possible realization of such a spin-nematic phase.
Theoretical investigations of the bilinear-biquadratic $S=1$ Heisenberg model on this lattice geometry have shown
that it is possible to stabilize spin-nematic LRO of ferroquadrupolar and antiferroquadrupolar type depending on the sign of the biquadratic
interaction,\cite{prl_triangularlattice,tsunetsugu_arikawa} and that applying an external
magnetic field leads to a remarkably rich phase diagram, with in particular a 2/3 magnetization plateau above
the antiferroquadrupolar phase.\cite{prl_triangularlattice}

In this paper we investigate the one-dimensional version of the model defined by the Hamiltonian
\begin{equation}
\mathcal{H} = J \sum_i \left[\cos \theta \, \mathbf{S}_i \cdot \mathbf{S}_{i+1} + \sin \theta \left(\mathbf{S}_i \cdot \mathbf{S}_{i+1}\right)^2\right] - H S^z_{\text{tot}},
\label{eq:ham}
\end{equation}
where $J,H>0$.  
While a variety of materials realizing $S=1$ Heisenberg systems are known,\cite{springer_quantummagnetism}
the biquadratic term seems to be more difficult to realize in nature.\cite{mila_zhang}
However, recent progress in the realization of effective spin-Hamiltonians in systems of ultracold atomic gases on optical lattices\cite{Cirac_PRL_spinhamiltonians_OptLatt,S.Trotzky01182008,sun_hamiltonians} opens up a promising alternative route
to investigate such systems in experiments.

At zero field, the properties of this model are well understood by now.
\cite{uimin,lai,sutherland,takhtajan,babujian,PRL_AKLT,barber1989,kluemper1989,Xian1993,fath1991,fath1993,Schollwock1996,buchta2005,PRB_Lauchli}
For our considerations, it is helpful to keep in mind the following aspects of its phase diagram.
Between the two integrable points $\theta_{\rm TB} = -\pi/4$ (the Takhtajan-Babujian point,\cite{takhtajan,babujian} TB) and $\theta_{\rm ULS} = \pi/4$ (the Uimin-Lai-Sutherland point,\cite{uimin,lai,sutherland} ULS) the chain possesses a finite Haldane gap.\cite{haldane1983,haldane1983PRL}
In this phase, at the so-called AKLT point (named after Affleck, Kennedy, Lieb, and Tasaki, Ref.~\onlinecite{PRL_AKLT}) $\theta_{\rm AKLT} = \arctan 1/3$, the ground state is exactly known
to be a valence bond solid, and for $\theta > \theta_{\rm AKLT}$ the spin correlation functions become incommensurate.\cite{Schollwock1996,fath_suto_2000}
Between the ULS point and $\theta = \pi/2$, the system is gapless and shows antiferroquadrupolar spin-nematic quasi long range order (QLRO).\cite{PRB_Lauchli}
For negative values of $\theta$, between $\theta = -3/4 \pi$ and $\theta = -\pi/4$ the system is dimerized and has a finite gap.\cite{buchta2005,barber1989,kluemper1989,Xian1993}
For the remaining values of $\theta$, the system is in a ferromagnetic state.

In comparison, the properties in a field have been investigated less intensively. Most of the attention has been devoted
to the region of positive and not too large biquadratic interaction,\cite{parkinson1989,okunishi1999rapid,okunishi1999,fath_littlewood}
and large regions of the phase diagram remain unexplored. For instance, the fate of the spin-nematic
phase realized at zero field in the region $\pi/4 < \theta < \pi/2$\cite{PRB_Lauchli} remains an open
issue, and the finite field properties of the model for negative biquadratic interactions are largely 
unexplored, including the finite field properties of the integrable TB point.

In this paper, we complete previous studies and consider the {\it full} phase diagram at finite magnetic fields over the whole range of $\theta$ by applying the density matrix renormalization group method (DMRG),\cite{white1992,schollwoeck2005} complemented by a Bethe ansatz (BA) solution of the TB and ULS points in a magnetic field and field theoretical considerations.
We put special emphasis on the possible realization of spin-nematic QLRO in the presence of a field by explicitly computing the spin-nematic correlation functions in real space. 
\begin{figure*}[t]
\includegraphics[width=0.495\textwidth]{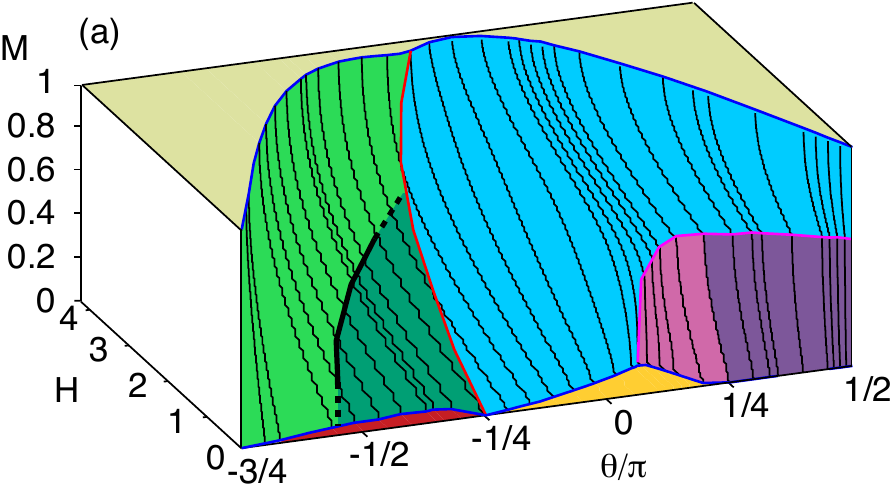}
\includegraphics[width=0.495\textwidth]{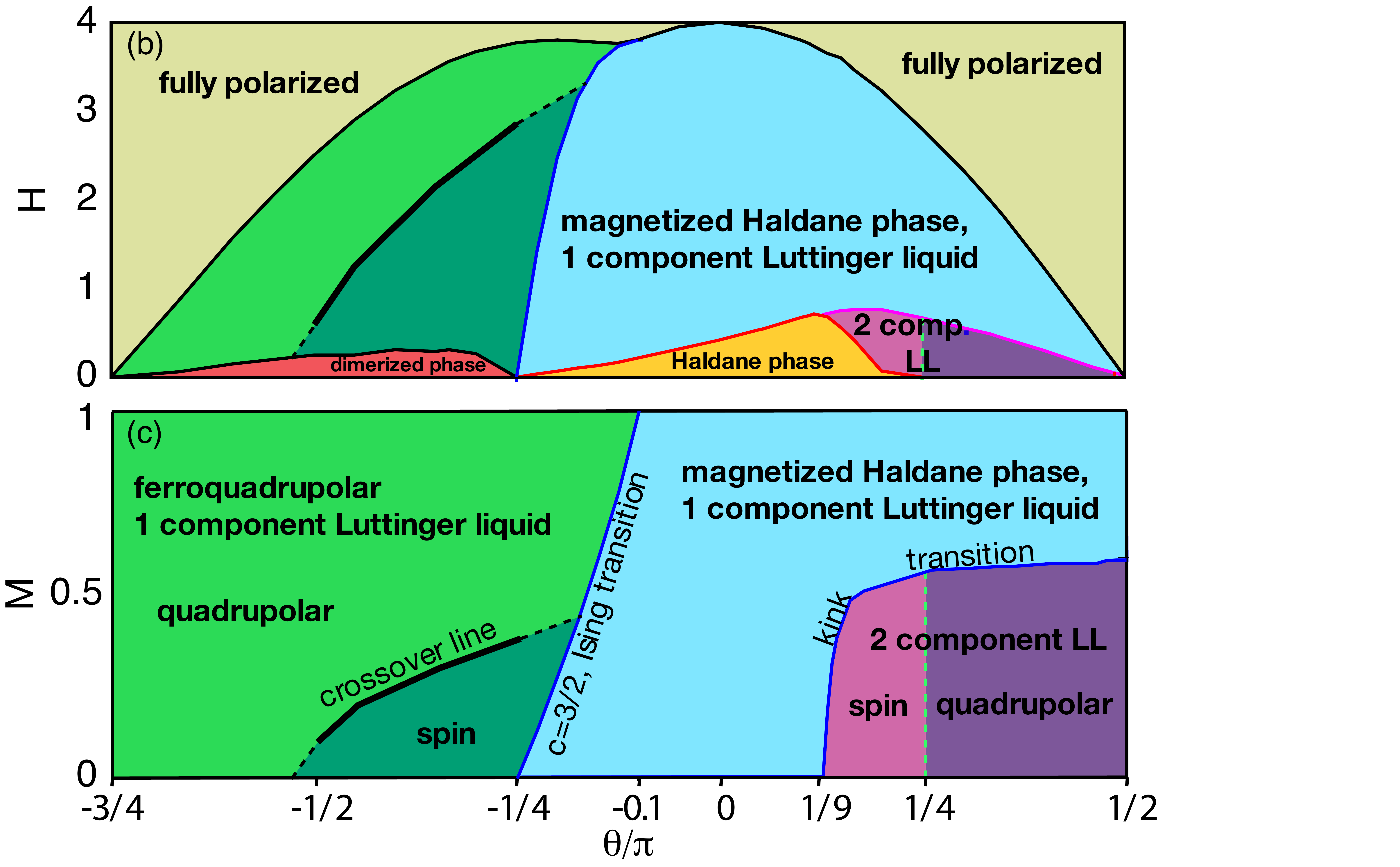}
\caption{(Color online) (a) DMRG results for systems with $L=60$ lattice sites for the magnetization of the BLBQ $S=1$ chain as a function of $(H,\theta)$. The various colors indicate the different phases of the system. (b) Phase diagram obtained from the DMRG results for the magnetization, the correlation functions defined in Eqs.~(\ref{eq:Slongcorr})-(\ref{eq:Qcorr}), (\ref{eq:chiralcorrs}) and for the central charge $c$ as a function of $(H,\theta)$ and (c) as a function of $(M,\theta)$.
The black line in the magnetized dimer phase indicates a crossover line at which the exponents of $C_S^{\rm long}$ and $C_{Q,2}$ are both equal to one, see Sec.~\ref{sec:crossovernegtheta} (the dashed lines are linear extrapolations to the boundary of the phase and serve as a guide to the eye). The green dashed line below the kink transition indicates a crossover line between two different two-channel LL phases, one of them being a spin-nematic LL. Note that, as discussed in Sec.~\ref{sec:crossoverpostheta}, this line is not exactly at $\theta = \pi/4$ but seems to wind around this value.}
\label{fig:finitefieldphasedia_chain}
\end{figure*}

The scope of this paper is fourfold.
First, by considering the full phase diagram of the system at finite magnetic fields we want to make precise predictions for ongoing efforts in the realization of such models in systems of ultracold atomic gases on optical lattices and for future quantum magnetic materials which eventually may be described in terms of the BLBQ chain.
Second, we address the possibility to realize unconventional QLRO by explicitly computing the spin-nematic correlation functions in real space. Third, we consider in detail the phase transitions at finite field and study their critical behavior.
Fourth, we address the possibility to realize vector chiral LRO as identified previously in frustrated $S=1/2$ chains in a magnetic field and has been proposed for the BLBQ chain in a magnetic field.\cite{kolezhuk_vekua2005}
We will demonstrate that the magnetic field leads to the realization of five different LL phases,
and that these magnetic phases are connected to each other by either continuous phase transitions or crossovers.
Our findings are summarized in Fig.~\ref{fig:finitefieldphasedia_chain} which show our DMRG results for the magnetization as a function of $(H,\theta)$ and the main result of this paper which is the complete phase diagram of the BLBQ chain in a magnetic field.

The paper is organized as follows.
In Sec.~\ref{sec:observables} we introduce the observables relevant for the description of the various LL phases.
In Sec.~\ref{sec:phasedia} we present the complete phase diagram as a function of $(H,\theta)$ by discussing our results for the Magnetization (Sec.~\ref{sec:magphasedia}), for the central charge (Sec.~\ref{sec:centralchargephasedia}), for the correlation functions in the single-component LL phases (Sec.~\ref{sec:ferroquadLL}), and for the correlation functions in the two-component LL phases (Sec.~\ref{sec:2LLphases}). 
Concerning the single-component LL phases, we demonstrate in Sec.~\ref{sec:ferroquadLLphase} that a ferroquadrupolar LL phase is realized, and we discuss the extension of the magnetized Haldane phase in Sec.~\ref{sec:magnetizedHaldane}.
In Sec.~\ref{sec:absencechiralorder} we demonstrate the absence of vector chiral order in the two-component LL phases, and in Sec.~\ref{sec:crossoverpostheta} we show that one of them is a spin-nematic LL.  
In Sec.~\ref{sec:pairing} we discuss in detail the transition from the magnetized dimer phase to the magnetized Haldane phase, which we identify to be an Ising transition with central charge $c=1+1/2=3/2$.
This scenario is further corroborated by a field theoretical treatment in the vicinity of the TB point discussed in Sec.~\ref{sec:fieldtheory}.
We summarize our findings and conclude in Sec.~\ref{sec:summary}.
Finally, we provide in Appendix~\ref{appendix} and~\ref{appendix2} a more detailed discussion of the BA solutions of the model at the TB and the ULS point at finite magnetic fields, respectively. 

\section{Observables and Correlation Functions}
\label{sec:observables}

\subsection{Magnetization}
\label{sec:magnetization}
The properties of the model have been identified by calculating with DMRG a number of
characteristic quantities. The first one is the magnetization defined by
\begin{equation}
M = \frac{1}{L} \sum\limits_i \langle S_i^z \rangle,
\label{eq:mag}
\end{equation}
which has been determined as a function of $\theta$ and applied magnetic field $H$ [Fig.~\ref{fig:finitefieldphasedia_chain}(a)].

\subsection{Spin correlation functions}
\label{sec:spincorrels}
The second source of information comes
from the behavior of correlation functions characterizing magnetic, spin-nematic and vector-chiral (quasi-)long-range order.
We investigate possible algebraic decay of these correlation functions and compare the numerical values of the exponents with each other, the exponent with the smallest absolute value giving the dominant correlation function. This
is of particular interest for the characterization of the gapless LL phases at finite field. 

The first type of QLRO is identified by the correlation functions of the local spins,
\begin{eqnarray}
C_{S}^{\text{long}}(i,j) & = & \langle S_i^z S_j^z \rangle - \langle S_i^z \rangle \langle S_j^z \rangle \label{eq:Slongcorr}\\
C_{S}^{\text{trans}}(i,j) & = & \langle S_i^- S_j^+ \rangle. \label{eq:Stranscorr}
\end{eqnarray}
In 1D, an algebraic decay of $C_S^{\text{long}}(i,j)$ indicates magnetic QLRO along the field, while a power-law behavior of $C_S^{\text{trans}}(i,j)$ can be interpreted as magnetic QLRO perpendicular to the field or as
a quasi-condensate of magnons.

\begin{figure}[b]
\includegraphics[width=0.45\textwidth]{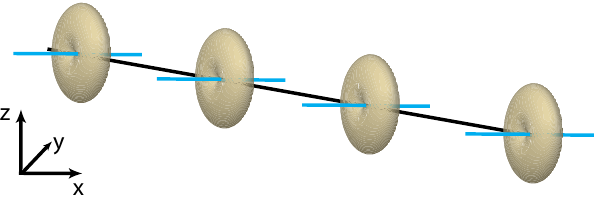}
\caption{(Color online) Sketch of a ferroquadrupolar ordered state in zero magnetic field.
Note that the breaking of the spin-SU(2) symmetry can be described by a director located on
each lattice site around which the spins fluctuate. The parallel alignement of
these directors perpendicular to the $z-$axis is representative for the ferroquadrupolar state.}
\label{fig:quadrupole}
\end{figure}

\subsection{Quadrupolar correlation functions}
\label{sec:quadrupolarcorrels}
For systems with $S>1/2$, however, QLRO can also be identified by considering the on-site spin-nematic correlation functions
\begin{equation}
C_Q(i,j) = \langle \vec{Q}_i \cdot \vec{Q}_j\rangle - \langle \vec{Q}_i \rangle \cdot \langle \vec{Q}_j \rangle,
\label{eq:Qcorr}
\end{equation}
where we have introduced the local quadrupolar order parameter
\begin{equation}
 \vec{Q}_i =
\left(
\begin{array}{c}
\frac{2}{\sqrt{3}}\left[\left( S_i^z \right)^2 - \frac{1}{4} \left(S_i^+ S_i^- + S_i^- S_i^+ \right) \right] \\ 
\frac{1}{2} \left(S_i^+ S_i^z + S_i^z S_i^+ + S_i^- S_i^z + S_i^z S_i^-\right) \\ 
-\frac{i}{2} \left(S_i^+ S_i^z + S_i^z S_i^+ - S_i^- S_i^z - S_i^z S_i^-\right)\\ 
-\frac{i}{2} \left[\left( S_i^+ \right)^2 - \left( S_i^- \right)^2 \right] \\ 
\frac{1}{2}\left[ (S_i^+)^2 + (S_i^-)^2 \right] 
\end{array}
\right).
\label{eq:Qvector}
\end{equation}
Note that only the first entry of this vector conserves $S^z_{\rm total}$, while the other ones change this quantum number by $\Delta S^z = 1$ or $\Delta S^z = 2$, respectively.
This leads to three components of the correlation functions (\ref{eq:Qcorr}): the longitudinal component $C_{Q,0}$ considering the terms conserving $S^z_{\rm total}$, the transverse component $C_{Q,1}$ considering the entries of (\ref{eq:Qvector}) with $\Delta S^z = \pm 1$, and the pairing component $C_{Q,2}$ considering the entries with $\Delta S^z = \pm 2$.
In particular, we compute
\begin{eqnarray}
 \mbox{\hspace{-5cm}} \qquad C_{Q,0}(i,j) &=&
 2 \left( \frac{1}{12} \left\langle
\left(S_i^- S_i^+ + S_i^+ S_i^- - 4\left(S_i^z\right)^2\right) \right. \right. \nonumber \\
&& \left.
\times \left(S_j^- S_j^+ + S_j^+ S_j^- - 4\left(S_j^z\right)^2\right)
\right\rangle  \nonumber \\
&& \left. - \left\langle Q_i^{(1)} \right\rangle \left\langle Q_j^{(1)}
\right\rangle \right), \label{eq:Q0}\\
 C_{Q,1}(i,j) &=&
\frac{1}{2} \left\langle
\left( S_i^+ S_i^z + S_i^z S_i^+ \right)
\left( S_j^z S_j^- + S_j^- S_j^z \right) +
h.c.
\right\rangle \label{eq:Q1}\\
\hspace{-5cm}C_{Q,2}(i,j) &=& \frac{1}{2} \left\langle
  \left(S_i^+\right)^2 \left(S_j^-\right)^2 + h. c. \right\rangle \label{eq:Q2}
\end{eqnarray}
At zero magnetic field, these three components are identical due to the SU(2) symmetry of the system.
At finite field, however, the SU(2) symmetry is reduced to U(1) and the different components can show different behavior and characterize different types of QLRO.
In addition, in the presence of a finite field, the operators entering
$C_{S}^{\text{long}}$ and $C_{Q,0}$ are allowed to mix by symmetry. These correlation functions are thus expected to
decay with the same power law and to test for the same type of QLRO, namely magnetic QLRO along the field.
Similarly, $C_{S}^{\text{trans}}$ and $C_{Q,1}$ can mix and test for magnetic QLRO transverse to the field,
or for magnon quasi-condensation.
However, the pairing component $C_{Q,2}$ has no magnetic partner and probes possible QLRO of non-magnetic type.
In the following we will refer to a phase at finite magnetizations in which this component decays algebraically
and dominates as a quadrupolar or {\it spin-nematic Luttinger liquid}.
In analogy to the interpretation of (\ref{eq:Stranscorr}), QLRO in $C_{Q,2}$ can also be viewed as the quasi-condensation of $S=2$ bound pairs of magnons.
If, in addition, the structure factor of this component of the quadrupolar correlation function is peaked at a wavevector $q=0$, we call the phase a {\it ferroquadrupolar Luttinger liquid} which we sketch in Fig.~\ref{fig:quadrupole}: on a single site, a finite expectation value $\langle \vec{Q}_i \rangle$ can be envisaged as fluctuations around an axis called {\it director}.\cite{prl_triangularlattice} In a ferroquadrupolar phase, all directors align parallel to each other. Note that the directors are perpendicular to the field.
In previous work, ferroquadrupolar long-range-order has been identified for the $S=1$ bilinear-biquadratic Heisenberg model on the square\cite{harada_kawashima} and triangular lattice\cite{prl_triangularlattice}
 at zero magnetic field. In the 1D case, at zero field spin-nematic phases have been identified numerically.\cite{PRB_Lauchli,prb_corboz} They are not of ferroquadrupolar type since the structure factor is peaked at some finite momentum.
However, in Sec.~\ref{sec:ferroquadLLphase} we demonstrate that such a phase is realized in 1D in the BLBQ chain in a magnetic field.
Note that beyond the quadrupolar order it is possible to realize multi-polar (quasi-)long-range order in spin systems by considering bond products of spin-operators as described in Refs.~\onlinecite{vekua2007,Kecke2007,Hikihara2008,PRB_Sudan}.
However, in the following we will restrict ourselves to the investigation of possible quadrupolar order.

\subsection{Vector chiral correlation functions}
\label{sec:vectorchiralcorrels}

In addition to possible QLRO associated to breaking the SU(2) or U(1) symmetry, true LRO can be obtained by breaking the parity of the system.
This is tested by the first of the following two vector chiral correlation functions,
\begin{eqnarray}
C_{\kappa}^{\text{long}}(i,j) &=& \langle \kappa^z_i  \kappa^z_j \rangle \nonumber \\
C_{\kappa}^{\text{trans}}(i,j) &=& \langle \kappa^x_i  \kappa^x_j \rangle  =  \langle \kappa^y_i  \kappa^y_j \rangle, \nonumber\\
\textrm{where } \vec{\kappa}_j &=&  \mathbf{S}_j \times \mathbf{S}_{j+1}, \label{eq:chiralcorrs}
\end{eqnarray}
Such vector chiral order has been, e.g., identified using the DMRG in frustrated $S=1/2$ and $S=1$ Heisenberg chains.\cite{IanMcCulloch_VectorChiralOrder,PRB_Sudan}
In this paper, we consider a proposal of Kolezhuk and Vekua\cite{kolezhuk_vekua2005} in which a vector chiral phase at finite magnetizations for positive values of $\theta$ has been suggested.
We address this issue by directly computing the correlation functions~(\ref{eq:chiralcorrs}).

\subsection{Correlation exponents}
\label{sec:correlexponents}

Whenever one of the above correlation functions decays as a power law at long distance,
the decay will be described by a positive exponent $\eta$ according to:
\begin{equation}
C(i,j) \propto \vert i-j \vert^{-\eta}.
\end{equation}
The exponents will be distinguished by the same indices and superscripts as the corresponding correlation functions:
$\eta_S^{\text{long}}$, $\eta_S^{\text{trans}}$, $\eta_{Q,0}$, $\eta_{Q,1}$, $\eta_{Q,2}$. If several Fourier
components of the correlation functions decay algebraically, the exponents will be distinguished by
an additional index.

As pointed out above, in a magnetic field some spin and quadrupolar correlation functions are coupled,
so that $\eta_S^{\text{long}}=\eta_{Q,0}$ and $\eta_S^{\text{trans}}=\eta_{Q,1}$. So we end up with
three a priori independent exponents: $\eta_S^{\text{long}}$, $\eta_S^{\text{trans}}$ and $\eta_{Q,2}$.

\subsection{Central charge}
\label{sec:centralcharge}

The analysis of the phase diagram is complemented by computing the central charge $c$.
For systems amenable to a description by conformal field theory this quantity characterizes the phase and the universality class of phase transitions.\cite{book_CFT}
Using the DMRG, it can be obtained easily by computing the von Neumann entanglement entropy of blocks of consecutive sites
\begin{equation}
S_\ell = - {\rm Tr} \varrho_\ell \ln \varrho_\ell,
\end{equation}
where $\varrho_\ell$ is the reduced density matrix of a subsystem of size $\ell$.
In order to circumvent the oscillations which appear in the case of open boundary conditions (see, Refs.~\onlinecite{prb_corboz,legeza_PRL} for the behavior at the TB and at the ULS points at zero field), we obtain $c$ from systems with periodic boundary conditions (PBC), for which
Calabrese and Cardy\cite{calabrese_cardy} have derived the general expression
\begin{equation}
S_\ell = \frac{c}{3}
 \ln\left(
 \frac{L}{\pi}
 \sin\left(
 \frac{\pi \ell}{L}
 \right)
 \right)
 + g_{\rm PBC}.
 \label{eq:S_PBC}
\end{equation}
The numerical value of $c$ is then obtained by computing $S_\ell$ for finite systems and fitting Eq. (\ref{eq:S_PBC}) to the results.

\section{Phase diagram of the BLBQ chain at finite magnetic fields}
\label{sec:phasedia}

\begin{figure}[t]
\includegraphics[width=0.45\textwidth]{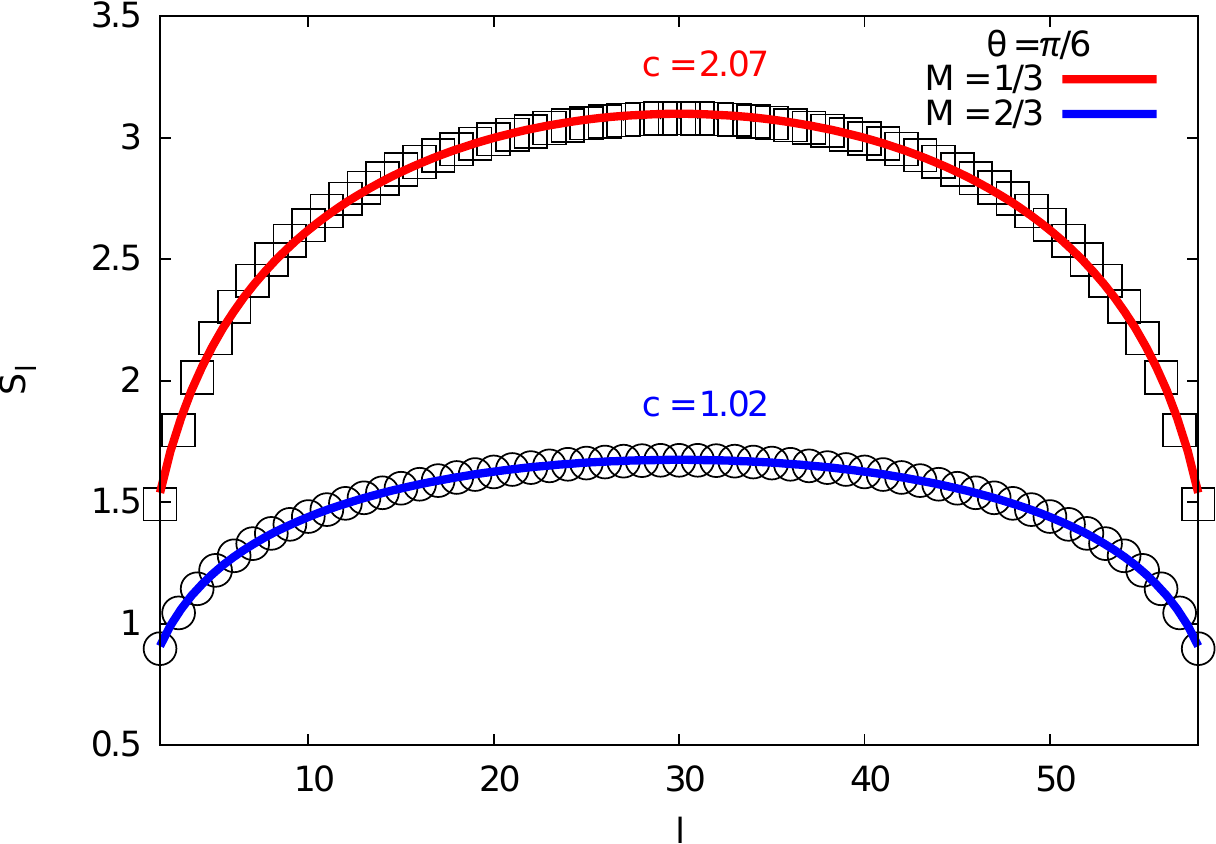}
\caption{(Color online) Block entropy for a system with $L=60$ lattice sites and PBC at $\theta=\pi/6$ below the kink transition ($M=1/3, \, \square$) and above ($M=2/3, \bigcirc$). The solid lines indicate the fit with Eq.~(\ref{eq:S_PBC}), the labels denote the central charge obtained by this fit.}
\label{fig:central_charge}
\end{figure}

In this section, we map out the phase diagram by successively looking at the magnetization, the central charge,
and the correlation functions.

\subsection{Magnetization}
\label{sec:magphasedia}

The magnetization is depicted in Fig.~\ref{fig:finitefieldphasedia_chain}(a).
Three separate regions of magnetization can be identified.
Starting from the dimerized phase,
the magnetization of the finite systems grows regularly in steps of $\Delta M = 2/L$, similar to the behavior
in certain polarized $S=1/2$ magnetic systems.\cite{PRB_Sudan,HeidrichMeisner2006,vekua2007,Kecke2007,Chubukov1991,Hikihara2008,HeidrichMeisner2009}. 
This is rather natural. Indeed, in this parameter range (large negative biquadratic interactions), the system dimerizes to form singlet pairs, and the
quintuplet crosses the triplet when the bilinear interaction vanishes.
In this region, the first excitation of
a pair of spins to cross the singlet upon applying a magnetic field is a quintuplet with $S^z=2$, leading to the step size of $2/L$.
In contrast to this, in an intermediate region around the Heisenberg point $\theta = 0$ the magnetization grows in steps of $\Delta M = 1/L$.
In the region $-\pi/4 \leq \theta \lesssim -\pi/10$ the results for finite systems indicate a line of transitions at which the magnetization steps change from $\Delta M =2/L$ to $\Delta M = 1/L$ upon increasing $\theta$ and keeping $H=const$, or upon increasing $H$ at constant $\theta$, indicating the presence of a phase transition. 
This phase transition will be discussed in detail in Sec.~\ref{sec:pairing}.
By further increasing $\theta$, we identify a kink anomaly developing at $\theta \gtrsim \pi/9$ and persisting up to $\theta = \pi/2$.
Its position on the magnetization curve grows quickly up to the ULS point and then seems to saturate at $M \approx 0.6$, in agreement with previous findings concentrating on the Haldane phase (Refs. \onlinecite{okunishi1999rapid,okunishi1999}) and the vicinity of the ULS point (Refs. \onlinecite{parkinson1989,fath_littlewood}).

\subsection{Central charge}
\label{sec:centralchargephasedia}
The central charge has been calculated throughout the phase diagram. It is equal to 2 below the kink, and equal
to 1 everywhere else (except along the transition line where the magnetization steps change from $\Delta M =2/L$ to $\Delta M = 1/L$, where it is equal to 3/2, see below). Typical results are plotted in Fig.~\ref{fig:central_charge}.

When the central charge is equal to 1, the system is a single component Luttinger liquid, and all correlation exponents are controlled by a single parameter, leading to specific relationships between the exponents 
of algebraic correlations. 

When the central charge is equal to 2, the system is a 2-component Luttinger liquid, and the
exponents of the algebraic correlations can be obtained from the dressed charge matrix (see appendix~\ref{appendix2}).

So magnetization and central charge reveal the presence of three main phases: two single-component Luttinger
liquid phases with magnetization steps $\Delta M =2/L$ and $\Delta M = 1/L$ respectively, and a two-component
Luttinger liquid phase. We now turn to a careful investigation of correlations inside these phases to identify
the nature of the dominant correlations.

\subsection{Correlation functions in the single component Luttinger liquid phases}
\label{sec:ferroquadLL}

\begin{figure}[b]
\includegraphics[width=0.4\textwidth]{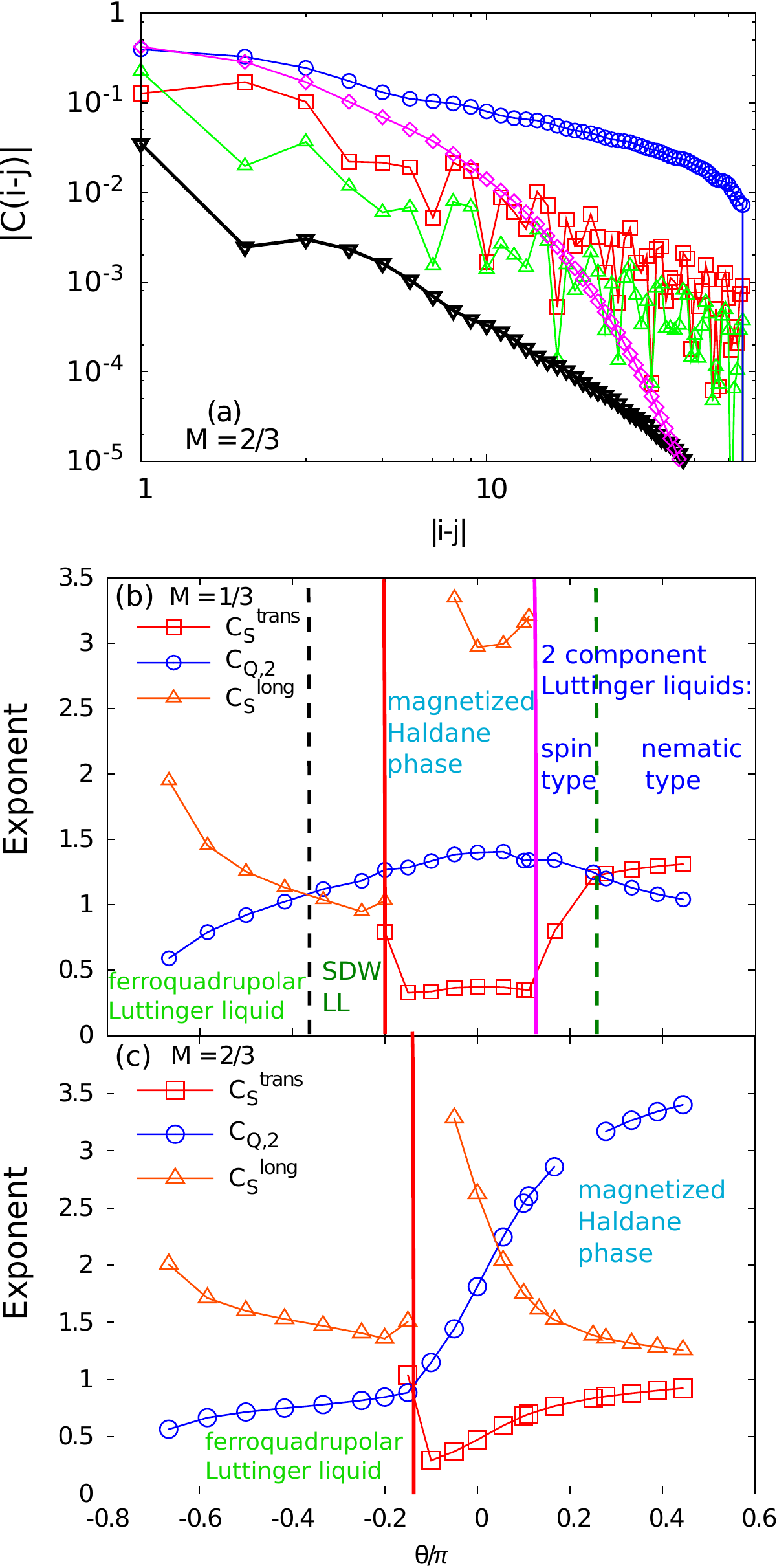}
\caption{(Color online) (a) Absolute value of the algebraically decaying correlation functions in the ferroquadrupolar LL phase at $M=2/3, \, \theta = -0.2 \pi$. 
The symbols and colors indicate $C_S^{\text{long}}$ [Eq.~(\ref{eq:Slongcorr}), red line, \textcolor{red}{$\square$}], $C_{Q,0}$ [Eq.~(\ref{eq:Q0}), green line, \textcolor{green}{$\triangle$}] and $C_{Q,2}$  [Eq.~(\ref{eq:Q2}), blue line, \textcolor{blue}{$\bigcirc$}], and $C_{\kappa}^{\text{long}}$  [Eq.~(\ref{eq:chiralcorrs}), black line, $\bigtriangledown$].
In addition, $C_S^{\text{trans}}$ [Eq.~(\ref{eq:Stranscorr}), magenta line, \textcolor{magenta}{$\diamond$}] - expected to decay exponentially -  is shown.
(b) Value of the exponents of $C_S^{\text{trans}}$ [Eq.~(\ref{eq:Stranscorr})], of $C_{Q,2}$ [Eq.~(\ref{eq:Q2})], and of $C_S^{\text{long}}$ [Eq.~(\ref{eq:Slongcorr})] as a function of $\theta$ at $M=1/3$. (c) The same as in (b) but at $M=2/3$. The vertical lines in (b) and (c) indicate the position of the phase transitions and crossovers as depicted in Fig.~\ref{fig:finitefieldphasedia_chain}.
Note that in the high field phase at $\theta = \pi/4$ the number of down spins is zero, leading to a vanishing $C_{Q,2}$.
In (b) and (c) the value of $\eta_S^{\text{long}}$ close to the phase transition at $\theta \approx -0.2 \pi$ is not shown since it is strongly affected by finite size effects. 
}
\label{fig:exponents}
\end{figure}

In this and the following section we base the characterization of the phases on the values of the exponents of $C_{Q,2}$, $C^{\rm trans}_S$ and $C^{\rm long}_S$.  
It is sufficient to consider only these correlation functions since, as expected due to the mixing at finite field, we find that transverse or longitudinal correlation functions decay with the same exponent, respectively.
An exception to this is the longitudinal component of the vector chiral correlation function, which we find to decay with an exponent of approx. 2 for all values of $M$ and $\theta$. 

The behavior of the exponents reveals important aspects of the phase diagram: 
As seen in Figs.~\ref{fig:exponents}(b) and (c), the transverse correlation functions decay algebraically with an exponent whose absolute value is much smaller than $\eta_{Q,2}$, as soon as the size of the magnetization steps changes to $\Delta M = 1/L$, revealing a fundamental change in the physics despite the fact that the central charge on both sides of the transition is $c=1$. 
The critical value of $\theta$ at which this transition takes place depends on the magnetization. 
This scenario is reminiscent of the paired superfluid phase and pair-unbinding transition to a superfluid of single bosons identified in Ref.~\onlinecite{schmidt2006} for systems of hard-core bosons with correlated hopping on the square lattice. At the present, it is unclear if in this system the transition is of first or second order. In the case of the BLBQ chain, however, we identify the transition to be a continuous one as we discuss in detail in Sec.~\ref{sec:pairing}. 
In the following, we address the various aspects concerning the correlation functions and the physics they reveal in more detail. 

\subsubsection{Exponent of the transverse correlation functions}
\label{sec:exponenttransvers}

In the discussion of the phase diagram, we mainly consider the exponents $\eta^{\rm trans}_S$ and $\eta_{Q,2}$.  
This is possible since we identify both, from the numerical data as well as from a field theory and Bethe ansatz approach (see Sec.~\ref{sec:fieldtheory}  and Appendix~\ref{appendix}), that the exponent $\eta^{\rm long}_S$ in the single channel LL phases is the inverse of the one of the corresponding transverse correlation function, i.e., $\eta^{\rm long}_S = 1/\eta_{Q,2}$ in the magnetized dimer phase and $\eta^{\rm long}_S = 1/\eta^{\rm trans}_S$ in the magnetized Haldane phase.  
Note, however, that the longitudinal correlation functions posses an additional oscillating component which makes it more difficult to obtain the numerical value of the exponent with a high precision. 
In order to obtain accurate results, we fit the exponent $\eta^{\rm long}_S$ to the Friedel oscillations in the local spin density $\langle S_i^z\rangle$ using expressions obtained by bosonization, as discussed in Sec.~\ref{sec:fieldtheory} below. 
We apply Eq. (\ref{friedelhaldane}) in the magnetized Haldane phase and Eq. (\ref{friedeldimer}) in the magnetized dimer phase by performing the fit only in the bulk region around the center of the system.
We find that in the magnetized Haldane phase close to the transition to the ferroquadrupolar LL, finite size effects become predominant due to the vicinity of the critical point. 
However, in contrast to the approach used in Ref.~\onlinecite{fath2003}, using these expressions it is possible to obtain accurate results without introducing additional phenomenological fitting parameters.
Note that in the region where $c=2$, there is no simple relation between the exponents of the various correlation functions, as discussed in more detail in Appendix~\ref{appendix2}. 
However, we find this exponent to be larger than $\eta^{\rm trans}_S$ and $\eta_{Q,2}$, so that we conclude that the longitudinal correlations do not become dominant in the two channel LL region.  

\subsubsection{Oscillatory component of the longitudinal correlation functions}
\label{sec:exponentoscillatory}
We find that in the magnetized Haldane phase the oscillatory component of the longitudinal correlations decays faster at larger magnetizations, and substantially faster than in the ferroquadrupolar LL phase. 
The frequency of this oscillation depends on the value of the magnetization, and changes upon crossing the phase transition.
It is remarkable that the wave vector of these oscillations in the ferroquadrupolar LL phase is $\pi (1-M)$, while at the Heisenberg point it is $2 \pi M$.\cite{Konik2002,FZ}
This can be understood in terms of the bound pairs of magnons populating the lattice, leading to an effective filling which is only half the value of the magnetization. 
This aspect will be discussed in more detail in Sec.~\ref{sec:fieldtheory} in the context of a field theoretical treatment. 

\begin{figure}[t]
\includegraphics[width=0.45\textwidth]{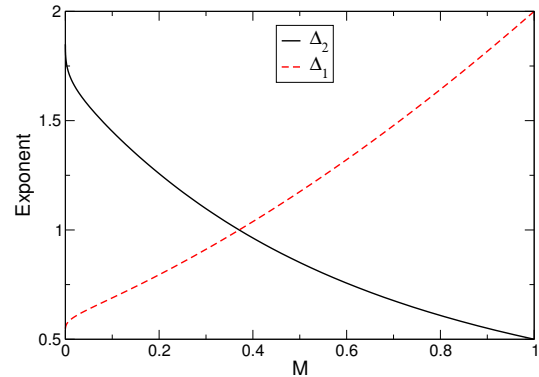}
\caption{(Color online) Exponents $\Delta_{1,2}$ of the power law decay of $C_S^{\rm long}$ and of $C_{Q,2}$, respectively, at the TB point ($\theta=-\pi/4$) as functions of the magnetization per site as obtained by Bethe ansatz (s. Appendix~\ref{appendix}). For small magnetizations $C_S^{\rm long}$ characterized by the exponent $\Delta_1$ dominates, while for large magnetization $C_{Q,2}$  characterized by the exponent $\Delta_2$ becomes dominant.}
\label{fig:Delta_M}
\end{figure}

\subsubsection{Ferroquadrupolar Luttinger liquid phase}
\label{sec:ferroquadLLphase}
We now focus on the region of the phase diagram in which the magnetization steps are of size $\Delta M = 2/L$ at high magnetizations (the light green region in Fig.~\ref{fig:finitefieldphasedia_chain}).     
In Fig.~\ref{fig:exponents}(a) we present our DMRG results for the various correlation functions as obtained for systems with $L=60$ lattice sites with open boundary conditions at $M=2/3$ and $\theta = -0.2 \pi$, a case which is representative for this phase.
The plot shows that the pairing component of $C_Q$ decays slowest, consistent with a spin-nematic phase.
Since the structure factor associated to this correlation function is peaked at a wave vector $q=0$, and the central charge is found to be $c=1$, we conclude that the system realizes a single
channel ferroquadrupolar LL phase.
At the same time, the transverse spin correlation function decays exponentially, showing that the single-spin excitation spectrum possesses a finite gap and that there is no quasi-condensation of magnons. This interpretation is further
confirmed by the analysis of the one-magnon and two-magnon spin gap, which is presented in more detail in Sec.~\ref{sec:pairing} below. 
In addition, a Bethe ansatz analysis at the integrable TB point confirms the numerical finding of a power law decay in $C_{Q,2}$ at finite field, while the one-magnon sector acquires a gap (see Appendix~\ref{appendix}). 

\subsubsection{Crossover from a ferroquadrupolar Luttinger liquid to a spin-density-wave Luttinger liquid}
\label{sec:crossovernegtheta}
In Fig.~\ref{fig:Delta_M} we show the exponents of the correlation functions $C_S^{\rm long}$ and $C_{Q,2}$ as a function of the magnetization as obtained by Bethe ansatz at the TB point $\theta_{\rm TB} = -\pi/4$. 
We observe that for magnetizations $M \lesssim 0.4$ the exponent of $C_S^{\rm long}$ is smaller than one, while for larger magnetizations $C_{Q,2}$ becomes dominant. 
This is in agreement with numerical results for these exponents.
Furthermore, we identify numerically for values of $\theta$ away from the integrable TB point the existence of a crossover line in the low field region which we show in Figs.~\ref{fig:finitefieldphasedia_chain}(b) and (c). 
At this crossover, the dominant correlations change from spin-nematic ones at large magnetizations to spin density wave (SDW) correlations. 
This is similar to a scenario realized by a frustrated ferromagnetic $S=1/2$ Heisenberg chain in a magnetic field, in which a crossover line divides spin-multipolar LL phases into a nematic and a SDW type of LL, as discussed in Refs.~\onlinecite{vekua2007,Hikihara2008,PRB_Sudan}.

\subsubsection{Magnetized Haldane phase}
\label{sec:magnetizedHaldane}
We find that the magnetized Haldane phase extends all over the region depicted in light blue in Fig.~\ref{fig:finitefieldphasedia_chain}, i.e., up to values of $\theta = \pi/2$ in the high field region above the kink transition. 
For the transverse spin correlations, the exponent can be obtained with reliable accuracy and at the Heisenberg point 
the obtained numerical value compares well with previously published results.\cite{Konik2002,Lorenzo2002,fath2003,prb_friedrich}
In the whole phase, the numerical data indicate $\eta^{\rm long}_S = 1/\eta^{\rm trans}_S$.
Interestingly, $\eta_{Q,2} \approx 4 \eta^{\rm trans}_S$, as can be seen in Figs.~\ref{fig:exponents}(b) and (c). 
Both findings are in agreement with predictions from field theory presented in Sec.~\ref{sec:fieldtheory}. 
The exponent of $C_Q$ at low magnetizations behaves rather smoothly at the transition. 
However, at larger magnetizations its magnitude increases quickly when crossing the phase transition, leading $C_{Q,2}$ to decay very fast in the high field region of the magnetized Haldane phase. 
This can be related to the small number of down spins in that region. 
The same effect is responsible for the complete suppression of these correlations at the ULS point due to the absence of down spins at this point which is discussed in Ref.~\onlinecite{fath_littlewood} and in Appendix~\ref{appendix}.

Note that $\eta^{\rm trans}_S$ jumps at the transition from the magnetized Haldane phase to the ferroquadrupolar LL phase. 
This is due to the nature of the phase transition and can be explained in terms of a field theory treatment of the transition, presented in Sec.~\ref{sec:fieldtheorytransition}. 

With this we conclude the discussion of the single component LL phases and turn now to the behavior of the correlation functions below the kink transition. 

\subsection{Correlation functions in the two-component Luttinger liquid phases}
\label{sec:2LLphases}

\begin{figure}[t]
\includegraphics[width=0.45\textwidth]{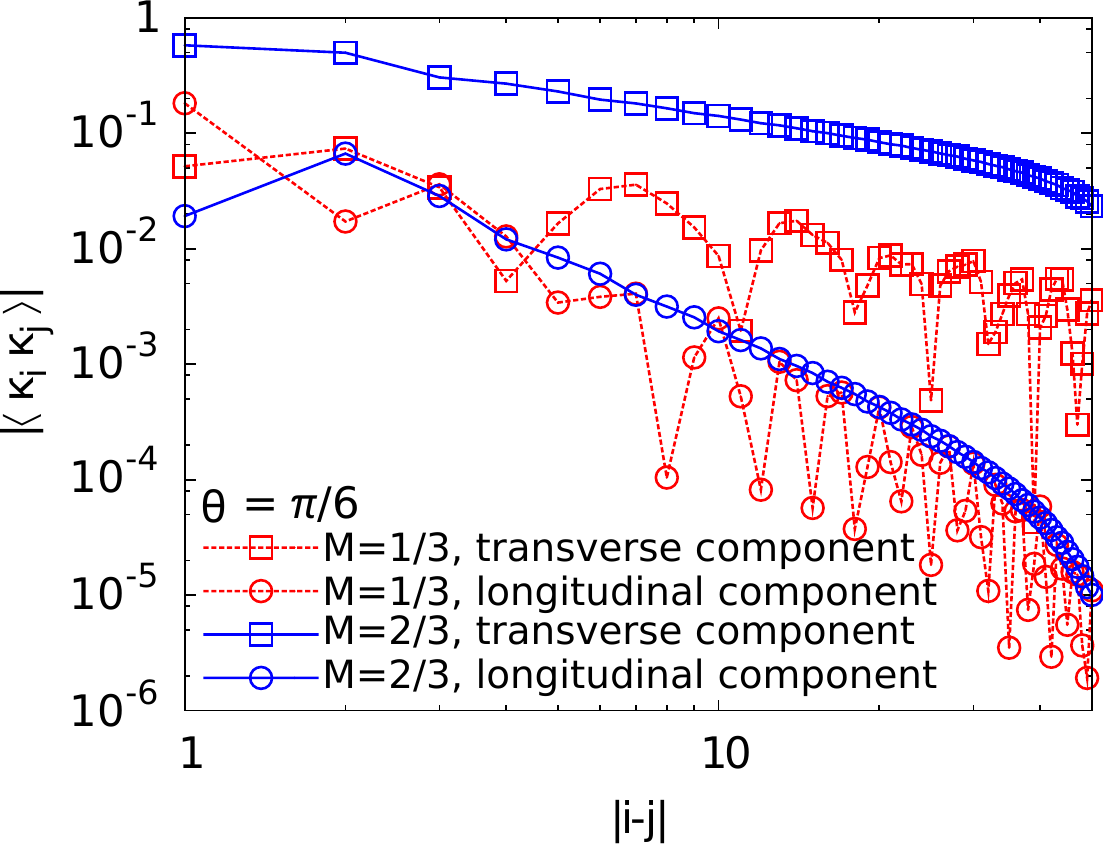}
\caption{(Color online) 
Vector-chiral correlation functions below ($M=1/3$, dashed lines) and above  ($M=2/3$, solid lines) the kink transition at $\theta = \pi/6$. 
}
\label{fig:chirals}
\end{figure}
\begin{table}[t]
\begin{tabular}{lccc}
& $k=0$ & $k_1$ & $k_2$\\
longitudinal correlations:  &&&\\
$C^S_{\rm long} $ & 1.91 & 1.75 & 1.77 \\ 
$C_{Q,0}$ & 1.75 & 1.58 & 1.71  \\
Bethe ansatz & 2 & 1.60 & 1.63   \\
\hline
transverse correlations:&&&\\
$C^S_{\rm trans} $ &  & 1.37  & 1.51\\ 
$C_{Q,1}$ &  & 1.38 & 1.51\\
Bethe ansatz &  & 1.18 & 1.28  \\
\hline
spin-nematic correlations: &&&\\
$C_{Q,2}$& 1.61 & 1.38 &  \\
Bethe ansatz & 1.68 & 1.20 & 
\end{tabular}
\caption{Comparison of the values for the exponents of the various correlation functions at the ULS point $\theta_{\rm ULS} = \pi/4$ at $M=1/3$ as obtained by DMRG and Bethe ansatz. We display the exponents of the non-oscillatory part, $k=0$, and the two smallest values at finite wave vectors $k_1$ and $k_2$.}
\label{tab:exponents}
\end{table}

In this section, we turn to the phases realized below the kink transition and characterize them by identifying the dominant correlation functions.  

\begin{figure}[b]
\includegraphics[width=0.48\textwidth]{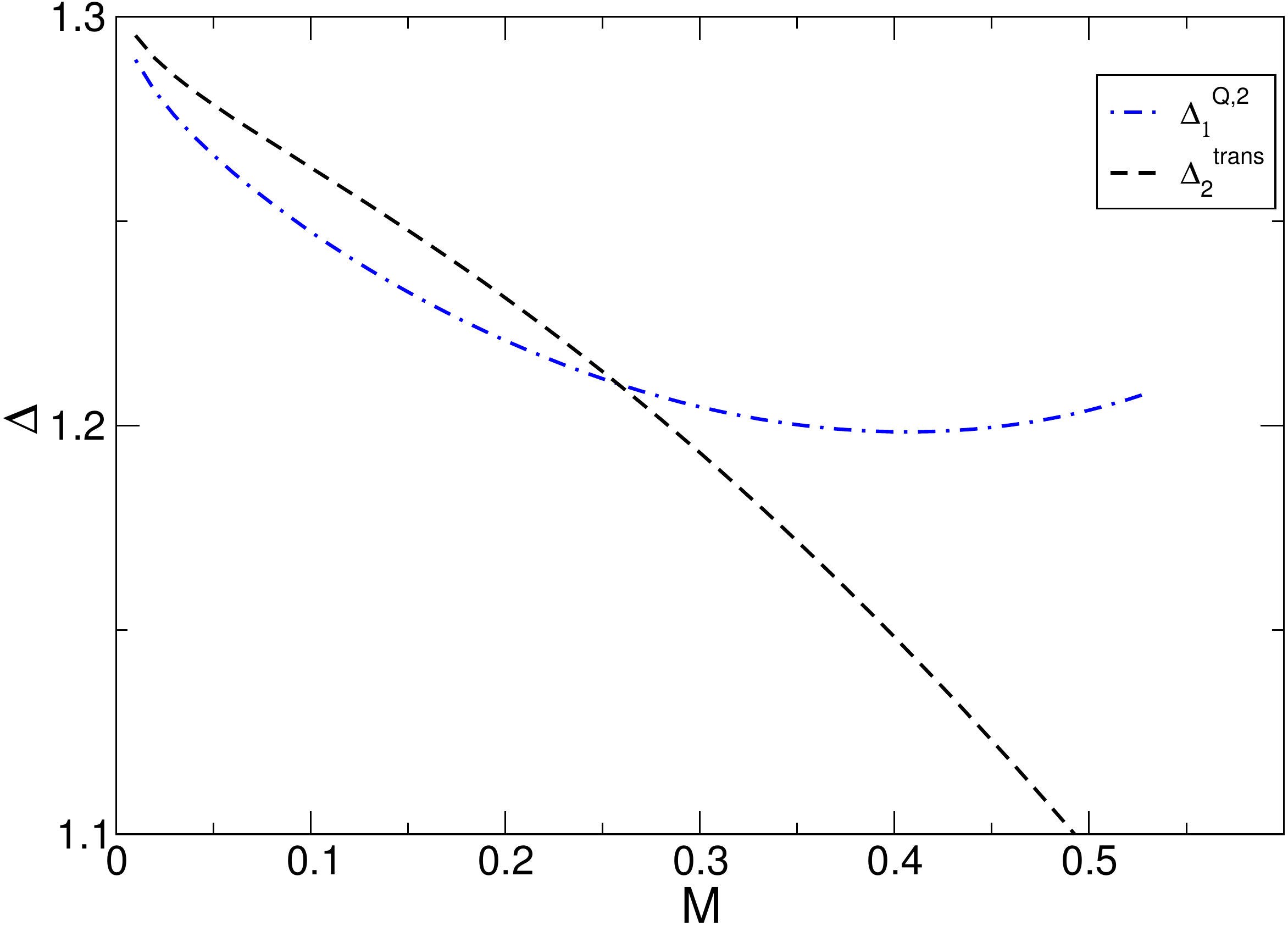}
\caption{(Color online) 
Bethe ansatz results for the 
exponents of the dominant component of the correlations at the ULS point $\theta_{\rm ULS} = \pi/4$ in the 2 LL region below the kink. 
The results shown are obtained by the Bethe ansatz calculation presented in Appendix~\ref{appendix2} and show the smallest values of the exponents of $C^{\rm trans}_S$ and $C_{Q,2}$ as a function of the magnetization per site. We observe a crossover between a regime where transverse spin correlations and quadrupolar correlations are dominant at $M \approx 0.258$. Note that the exponents of the other components of the correlations are larger and are discussed in more detail in Appendix~\ref{appendix2}.
}
\label{fig:su3exp2}
\end{figure}

\subsubsection{Absence of vector chiral order}
\label{sec:absencechiralorder}

The search for a parity broken phase is motivated by findings for frustrated $S=1/2$ chains, in which a kink transition separates two single channel LLs, one of them exhibiting vector chiral order.\cite{IanMcCulloch_VectorChiralOrder} 
By analogy, it has been suggested that the same scenario might occur in the present bilinear-biquadratic $S=1$ chain.\cite{kolezhuk_vekua2005}
However, as discussed in Sec.~\ref{sec:centralchargephasedia}, in the whole region below the kink the central charge is $c=2$, supporting a scenario in which two-component LL physics without vector chiral order is realized. 
This is confirmed by our results for the vector chiral correlation functions shown in Fig.~\ref{fig:chirals} which do not indicate the presence of parity breaking. 
We therefore conclude that vector chiral order is not realized and that below the kink transition in the whole region two channel LLs are realized which we will further characterize in the next section.

\subsubsection{Crossover from a SDW to a spin-nematic two channel Luttinger liquid}
\label{sec:crossoverpostheta}

As shown in Fig.~\ref{fig:exponents}(b), in the region $\pi/9 \lesssim \theta \lesssim \pi/4$ the transverse spin correlations are more dominant than the quadrupolar ones, while in the region $\pi/4 < \theta < \pi/2$ the quadrupolar correlations tend to be dominant. 
The presence of such a crossover line is confirmed by a Bethe ansatz analysis at the ULS point $\theta_{\rm ULS} = \pi/4$ presented in more detail in Appendix \ref{appendix2}. 
Note that the presence of the two massless modes leads to oscillating components at various momenta, which makes it difficult to obtain the numerical values of the exponents of the correlation functions. 
However, at the ULS point we can compare to the Bethe ansatz results, see Tab.~\ref{tab:exponents}. 
The numerical values are obtained for $M=1/3$, and a good agreement between the DMRG and the Bethe ansatz results is obtained. 

As shown in Fig.~\ref{fig:su3exp2}, the Bethe ansatz results demonstrate that for $M \lesssim 0.258$ the spin-nematic correlations are dominant, while at larger fields the transverse spin correlations are. 
A further numerical analysis of the exponents around $\theta = \pi/4$ indicates that the crossover line is, indeed, bent. It seems to exist at values $\theta < \pi/4$ for $M < 0.25$, while for $M > 0.25$ it seems to exist at $\theta > \pi/4$, but bending back towards $\pi/4$ upon further increasing $M$. 
However, since the values of the exponents are so close, it is very difficult to identify the exact position of this crossover line numerically and therefore we leave this aspect for future investigations. 
Due to this crossover line, the system seems to reflect to some extent the behavior at zero field, where the system is in the gapped Haldane phase for $\theta < \pi/4$, but realizes an antiferroquadrupolar $c=2$ critical state for $\theta > \pi/4$~\cite{PRB_Lauchli}. 
Note that this spin-nematic LL is not a ferroquadrupolar LL since the structure factor of the quadrupolar correlations is not peaked at a wave vector $q=0$, but at wave vectors which depend both on the value of $\theta$ and on the strength of the magnetic field $H$ as discussed in Ref.~\onlinecite{fath_littlewood} and in Appendix~\ref{appendix2}. 

Summarizing all these findings, we obtain the magnetic phase diagram of the BLBQ $S=1$ Heisenberg chain presented in Figs.~\ref{fig:finitefieldphasedia_chain}(b) and (c). 
In the next section, we will discuss in detail the nature of the transition from the ferroquadrupolar LL to the magnetized Haldane phase at finite magnetizations.

\section{Pair-unbinding transition}
\label{sec:pairing}

\begin{figure}[t]
\includegraphics[width=0.4\textwidth]{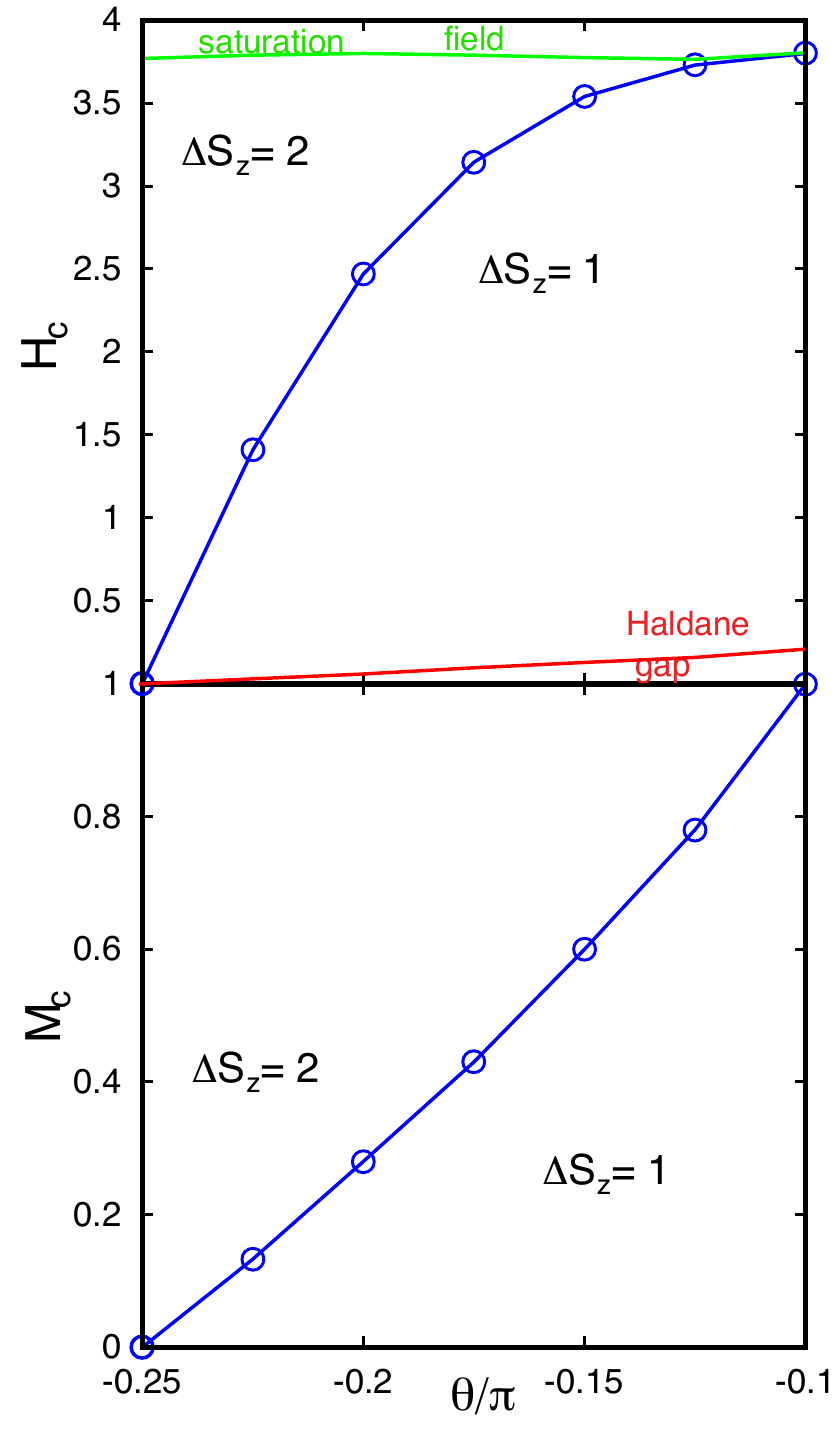}
\caption{(Color online) Line of phase transitions between the ferroquadrupolar LL and the magnetized Haldane phase (a) as a function of $(H,\theta)$ and (b) as a function of $(M,\theta)$. The data points indicate the position at which the step size of the magnetization changes from $\Delta M = 2/L$ to $\Delta M = 1/L$ for systems with $L=120$ lattice sites.}
\label{fig:pairingtransition}
\end{figure}

\begin{figure}[b]
\includegraphics[width=0.45\textwidth]{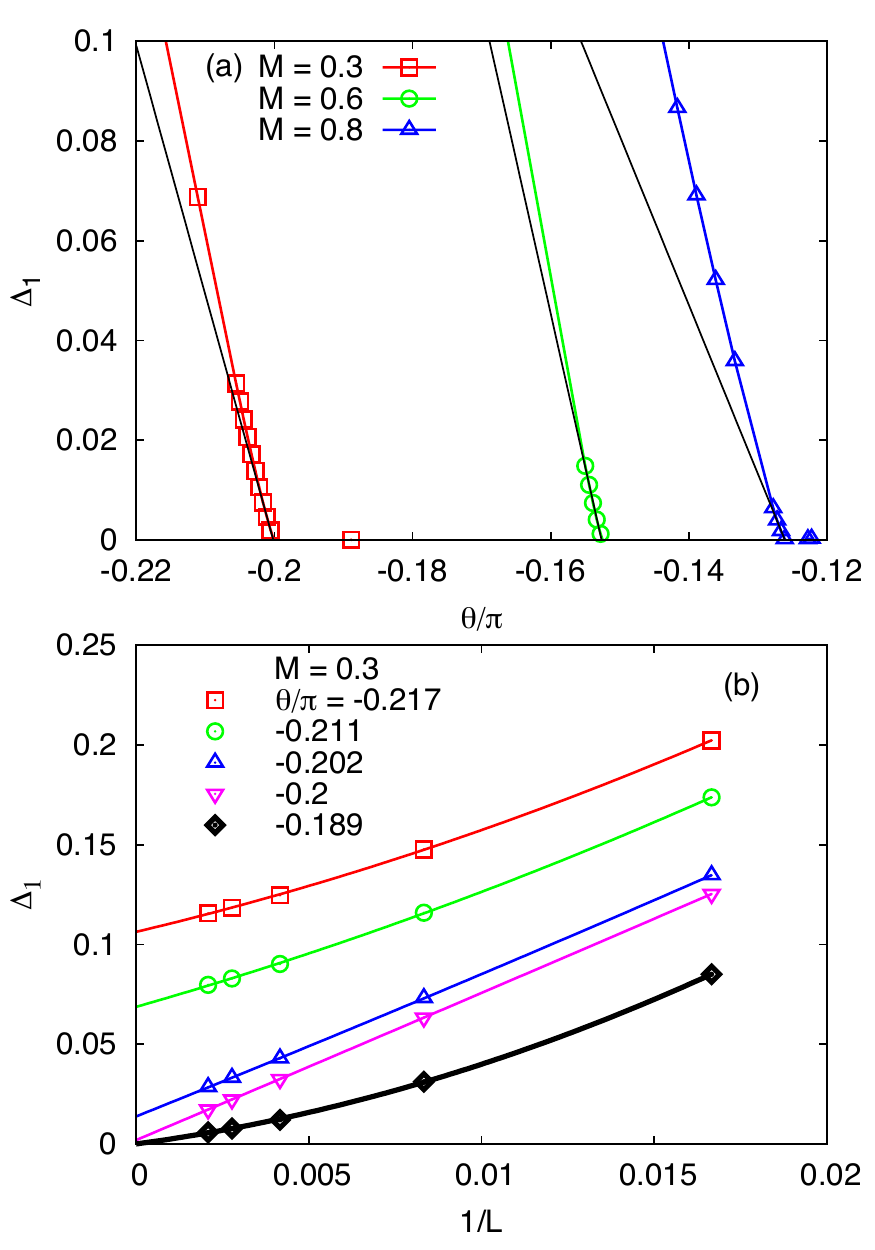}
\caption{(Color online) (a) $\Delta_1(\theta)$ at $M=0.3$, $M=0.6$ and $M=0.8$ after extrapolation to the thermodynamic limit. The black lines are linear fits in the vicinity of the critical points. (b) Example of the finite size scaling of the gap as a function of $1/L$ at $M=0.3$. The solid lines are fits with second order polynomials and are shown as a guide to the eye.}
\label{fig:gapnegativetheta}
\end{figure}

In this section, we focus on the nature of the phase transition between the ferroquadrupolar LL and the magnetized Haldane phase.
By considering systems with up to $L=480$ lattice sites, we do not find any indication for the formation of a jump in the magnetization curve, showing that a continuous rather than a first order transition (metamagnetic transition) is taking place.
Our findings for the magnetization curves indicate that an infinitesimal magnetic field at the TB point leads immediately to the binding of two magnons, while for $\theta > \theta_{\rm TB}$ closing of the Haldane gap leads to the condensation of single magnons.
On the other hand, for the fully polarized state, decreasing the field will lead to such bound states in the region $\theta \le \arctan(-1/3)$,\cite{karlo_communication} while for larger values of $\theta$ the excitations of the fully polarized state are described by single magnons.
We therefore conclude that the line of transitions is located between the TB point at zero field and $-\theta_{\rm AKLT}$ at the saturation field, as shown in more detail in Fig.~\ref{fig:pairingtransition}.
In the following we provide further support that this transition is a continuous one.
We consider the gaps
\begin{eqnarray}
\Delta_1 &=& \frac{1}{2} \left[E_0(S_z+1) + E_0(S_z - 1) - 2 E_0(S_z)\right]\\
\Delta_2 &=& \frac{1}{2} \left[E_0(S_z+2) + E_0(S_z - 2) - 2 E_0(S_z)\right]
\end{eqnarray}
which have been applied in Ref.~\onlinecite{vekua2007} for characterizing frustrated ferromagnetic $S=1/2$ chains at finite magnetizations.
$\Delta_1$ is a measure for the energy of single-spin excitations, while $\Delta_2$ correspondingly characterizes two-spin excitations.
If $\Delta_2 < \Delta_1$, then the lowest lying excitations are characterized as pairs of spins.
We find that, after extrapolating to the thermodynamic limit, $\Delta_2$ is always zero for all values of the magnetization in this parameter region.
This is confirmed by the observed algebraic decay of the quadrupolar correlation functions all over the parameter range.
However, $\Delta_1$ is finite in the upper part of the magnetization curves, but becomes zero at the point where the step size of the magnetization curves changes, and remains zero in the magnetized Haldane phase.
In Fig.~\ref{fig:gapnegativetheta} we show results for $\Delta_1(\theta)$ after extrapolating to the thermodynamic limit at three different values of the magnetization $M = 0.3$,  $M = 0.6$ and $M = 0.8$.
In all three cases, to a good approximation the gap in the vicinity of the transition point closes linearly and remains zero after the transition.
This supports the scenario of a line of continuous phase transitions.

\begin{figure}[b]
\includegraphics[width=0.45\textwidth]{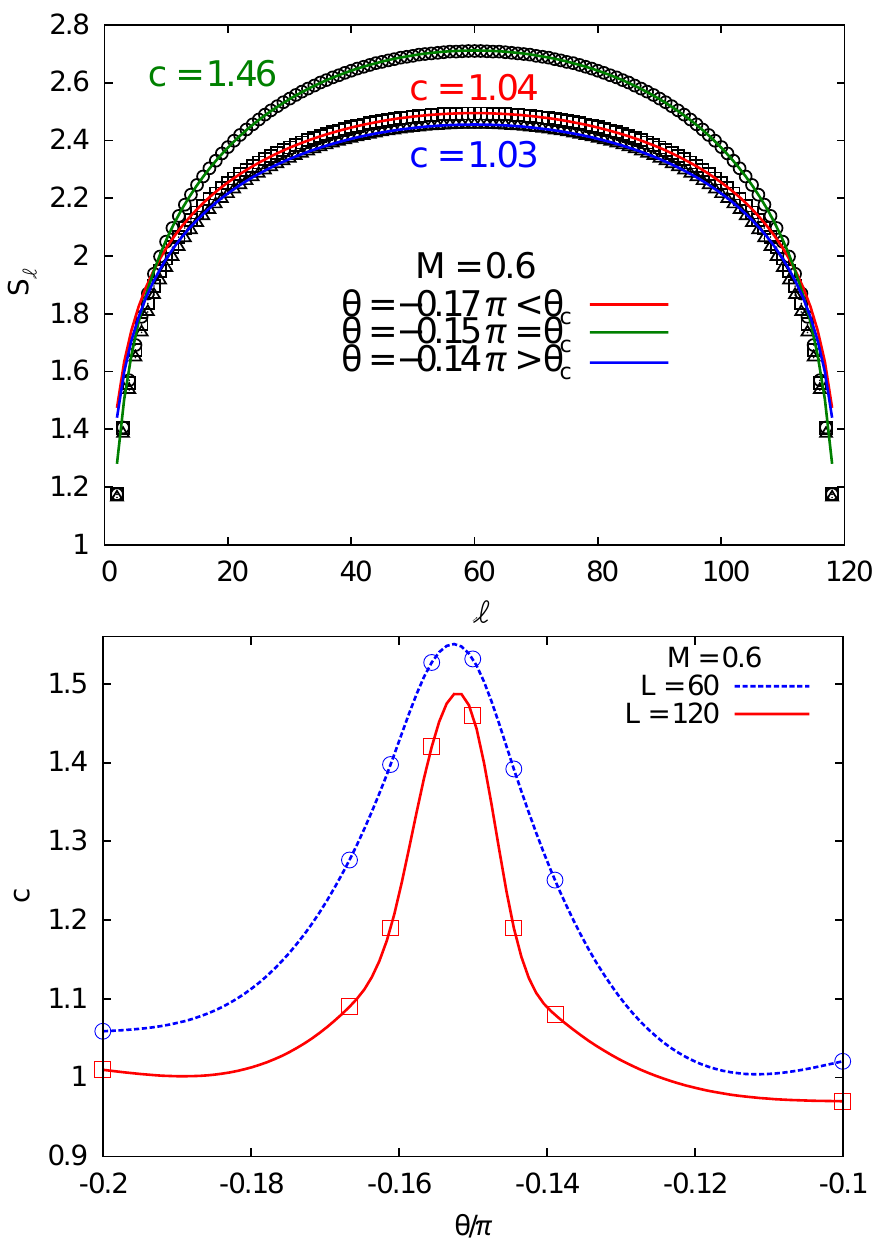}
\caption{(Color online) (a) Block entropy as obtained for systems with $L = 120$ lattice sites with PBC at $\theta \approx -0.17 \pi < \theta_c$ (squares), $\theta \approx -0.15 \pi \approx \theta_c$ (circles) and $\theta \approx -0.14 \pi > \theta_c$ (triangles) 
at $M=0.6$.
The value of the central charge as obtained by fitting with Eq.~(\ref{eq:S_PBC}) is shown.
(b) Central charge as a function of $\theta$ for systems with $L=60$ and $L=120$ sites with PBC. The data points are connected by a spline-interpolation as a guide to the eye. In the thermodynamic limit the maximum will be a sharp peak located at the critical point.}
\label{fig:entropies333}
\end{figure}

Next we address the universality class of the transition by computing the central charge $c$ on the line of phase transitions and in the adjacent phases.
In Fig.~\ref{fig:entropies333} we show our results for the block entropy for systems with PBC and $L=60$ or $L=120$ lattice sites keeping up to $m=2000$ density matrix eigenstates.
The value of the central charge is obtained by applying Eq.~(\ref{eq:S_PBC}) for different values of $\theta$ at fixed magnetization $M = 0.6$.
At the critical point, we obtain a value close to $c=3/2$, which is reproduced everywhere on the critical line, while we find the expected $c=1$ in both the ferroquadrupolar LL phase and the magnetized Haldane phase.

It is remarkable that the measured effective central charge at the transition in finite field is the same as the one at zero field at the TB point.
In this case the transition belongs to the $SU(2)_{k=2}$ Wess-Zumino-Witten-Novikov (WZWN) class,\cite{tsvelik_book}
with a unique set of scaling dimensions (and therefore exponents of the correlation functions).

\begin{figure}[t]
\includegraphics[width=0.45\textwidth]{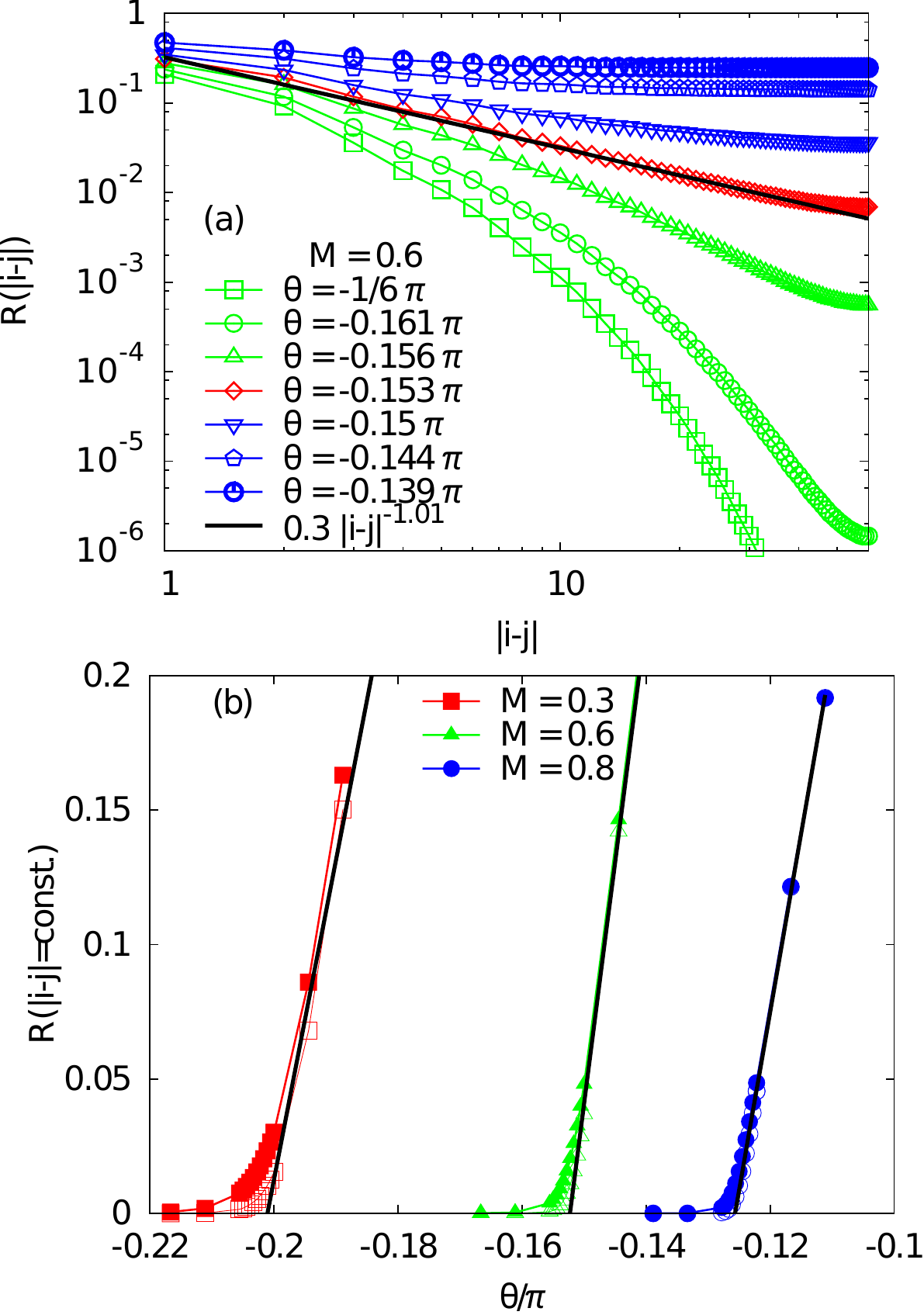}
\caption{(Color online) (a) Ratio $R(|i-j|)$ [Eq.~(\ref{eq:ratio})] as a function of $\theta$ at $M=0.6$. (b) Value of $R(30)$ (thick lines, filled symbols) and $R(50)$ (thin lines, empty symbols) as a function of $\theta$ at $M=0.3$, $M=0.6$ and $M=0.8$ obtained for systems with $L=120$ lattice sites and PBC. The black lines are linear fits in the vicinity of the critical points.}
\label{fig:Isingorderparameter}
\end{figure}
The nature of the transition in a finite field is discussed in some detail
using field theory methods in section \ref{sec:fieldtheory}. In
particular it is shown there that the value $c=3/2$ for the central charge
arises from an Ising degree of freedom that becomes critical on top of
an already present critical Luttinger liquid, i.e. $3/2=1+1/2$. 
This
is analogous to what has been found in
Ref. \onlinecite{Ejima} for a transition between two superfluids in a
two-component bosonic Feshbach problem,  
and in
Refs.~\onlinecite{lecheminant_prl,wu_prl,lecheminant_nuclphysb,capponi_rapid,roux2009} 
for the case of a fermionic $S=3/2$ Hubbard model.
In the attractive
$S=3/2$ Hubbard model the Ising transition separates a
quasi-condensate of pairs and a quasi-condensate of quartets of
fermions. 
This is reminiscent of our findings for the BLBQ chain in
which the phase transition is connecting a quasi-condensate of magnons
in the magnetized Haldane phase with the quasi-condensate of pairs of
magnons in the ferroquadrupolar LL phase, relating a single magnon in
the BLBQ chain to pairs of fermions in the $S=3/2$ Hubbard model. 

It would be interesting to identify the $Z_2$ symmetry of the Hamiltonian which gets broken at such pair-unbinding transitions. In higher dimensions, the U(1) symmetry of the system can be broken down partially to $Z_2$, and this remaining discrete symmetry can further spontaneously break at the phase transition (see, e.g., Ref.~\onlinecite{bonnes2011}). However, in one dimension, the continuous U(1) symmetry cannot be broken, and identifying the $Z_2$ symmetry is a more difficult task. This goes beyond the scope of the present paper in which we focus on presenting evidence in favor of such a scenario in the BLBQ chain at finite fields, and we leave a further characterization of this aspect open for future research.  

The Ising transition encountered in the models mentioned above can be characterized by considering 
particular {\it ratios} of correlation functions. \cite{lecheminant_prl} It is shown
in section \ref{sec:fieldtheory} that
\begin{equation}
R(|i-j|) = \frac{\langle S_i^- S_j^+ \rangle^4}{C_{Q,2}(i,j)}
\label{eq:ratio}
\end{equation}
can be used as a diagnostic of the Ising transition. More precisely,
the ratio $R(|i-j|)$ is related to a two-point function of an Ising
disorder field $\mu(x)$ by 
\begin{equation}
R(|i-j|) \propto= \langle \mu(x)\mu(0) \rangle^4.
\label{eq:ratio2}
\end{equation}
The magnetized Haldane phase corresponds to the disordered phase of
the Ising model, so that $\langle \mu(x)\rangle\neq 0$ and hence
$R$ is expected to tend to a finite value. On the other hand, 
the ferroquadrupolar LL phase corresponds to the ordered Ising phase
and hence $R$ will decay exponentially. At the transition itself the
field theory predicts $R(|i-j|) \sim 1/|i-j|$.

In Fig.~\ref{fig:Isingorderparameter}(a), we show our results for $R$
at fixed value of the magnetization when changing $\theta$. In the
ferroquadrupolar LL phase, this quantity decays exponentially, while
in the magnetized Haldane phase it indeed tends to a finite value.
At criticality, it decays $\sim 1/|i-j|$.
Note that in this case the value of one or both exponents of the
correlation functions needs to jump at the transition. As we will see
in Sec.~\ref{sec:fieldtheory}, it is the exponent of $C_S^{\rm trans}$
which behaves discontinuously, in agreement with the results shown in
Figs.~\ref{fig:exponents}(b) and (c). 
In Fig.~\ref{fig:Isingorderparameter}(b), we show the value of
$R(|i-j| = 30)$ and $R(|i-j| = 50)$ as a function of $\theta$ at
various values of the magnetization. Even though we are not
considering the limit of infinite separation between $i$ and $j$, the
behavior at the critical point is linear to a good approximation. 

Finally, we may consider the scaling of the Ising order parameter 
$\langle \mu(0)\rangle$ as a function of the deviation
$\theta-\theta_c$ from the critical point. As is shown in the next
section, this is related to $R$ by
\begin{equation}
\lim_{|i-j| \to \infty} R(|i-j|,\theta) \propto \langle\mu(0)\rangle^8
\propto |\theta-\theta_c|.
\end{equation}
Hence, the linear behavior shown in
Fig.~\ref{fig:Isingorderparameter}(b) is also in agreement with an
Ising transition.  Additional support for this scenario is given by
comparing the values of the critical points obtained by a linear
extrapolation of $R(30)$ and by a linear fit to $\Delta_1(\theta)$. 
From both sets of data, at $M=0.3$ we obtain $\theta_c \approx -0.2$,
at $M=0.6$ we obtain $\theta_c \approx  -0.15$, and at $M=0.8$ we find
$\theta_c \approx - 0.125$. Hence we conclude that starting from the
TB point at zero field, on the emerging line of phase transitions the
value of $c=3/2$ is kept, but the universality class changes from
~$SU(2)_{k=2}$ WZWN to Luttinger liquid plus Ising. This agrees with
the picture emerging from the field theoretical treatment of this
transition which we present in the next section. 

\section{Field Theory in the Vicinity of the Ising Transition}
\label{sec:fieldtheory}

Following Tsvelik \cite{amt} we can construct a field theory
description of the model in the vicinity of the Takhtajan-Babujian
point $\theta_{TB}=-\frac{\pi}{4}$, $H=0$. This results in a Hamiltonian of
the form
\bea
{\cal H}&=&\frac{iv}{2}\sum_{a=1}^3L_a\partial_xL_a-R_a\partial_xR_a
-im\sum_{a=1}^3R_aL_a\nn
&&+iH [L_1L_2+R_1R_2]+g\sum_{a=1}^3J^aJ^a\ ,
\label{HFT}
\eea
where $L_a$ and $R_a$ are left and right moving Majorana fermions and
\be
J^a=-\frac{i}{2}\epsilon^{abc}[L_bL_c+R_bR_c].
\ee
The mass $m$ in \fr{HFT} is proportional to the deviation $\theta-\theta_{TB}$
from the Takhtajan-Babujian point. The lattice spin operators are
expressed in terms of continuum fields as
\be
S^a_j\sim M^a(x)+(-1)^j n^a(x)\ ,
\ee
where $x=ja_0$ ($a_0$ is the lattice spacing). Here $M^a$
are related to the currents $M^a(x)\propto J^a(x)$, while
$n^a(x)$ are expressed in terms of the Ising order and
disorder operators as
\bea
n^x(x)\propto\sigma^1(x)\mu^2(x)\mu^3(x)\ ,\nn
n^y(x)\propto\mu^1(x)\sigma^2(x)\mu^3(x)\ ,\nn
n^z(x)\propto\mu^1(x)\mu^2(x)\sigma^3(x)\ .
\label{staggered}
\eea
We may bosonize the Majoranas 1 and
2 using
\bea
i[R_1L_1+R_2L_2]&\sim&\frac{1}{\pi\alpha}\cos\sqrt{4\pi}(\varphi_R+\varphi_L)
\ ,\nn
i[L_1L_2+R_1R_2]&\sim&\frac{1}{\sqrt{\pi}}\partial_x(\varphi_R+\varphi_L)\ ,\nn
L_1+iL_2&\sim&\frac{1}{\sqrt{\pi\alpha}}e^{-i\sqrt{4\pi}\varphi_L}\ ,\nn
R_1+iR_2&\sim&\frac{1}{\sqrt{\pi\alpha}}e^{i\sqrt{4\pi}\varphi_R}\ .
\eea
Here $\alpha$ is a short-distance cutoff.
Rewriting the Hamiltonian \fr{HFT} in terms of the canonical Bose
field $\Phi=\varphi_L+\varphi_R$ and the dual field
$\Theta=\varphi_L-\varphi_R$ results in
\bea
{\cal H}&=&{\cal H}_3+{\cal H}_B+{\cal H}_{\rm int}\ ,\nn
{\cal H}_3&=&\frac{iv}{2}\left[L_3\partial_xL_3-R_3\partial_xR_3\right]
-im\ R_3L_3\ ,\nn
{\cal
  H}_B&=&\frac{v'}{2}\left[\frac{1}{K}(\partial_x\Phi)^2+K(\partial_x\Theta)^2\right]
-\frac{m}{\pi\alpha}\cos\sqrt{4\pi}\Phi\nn
&&+\frac{H}{\sqrt{\pi}}\partial_x\Phi\ ,\nn
{\cal H}_{\rm int}&=&\frac{2ig}{\pi\alpha}\cos\sqrt{4\pi}\Phi\ L_3R_3.
\label{HFT2}
\eea

Here $K$ is a function of the applied magnetic field $H$ as well as the
parameter $\theta$. In order to deduce the structure of the ground
state phase diagram of \fr{HFT2} we may neglect the marginal term
${\cal H}_{\rm int}$. This leaves us with a decoupled theory of an
off-critical Ising model ${\cal H}_3$ and a sine-Gordon model with a
chemical potential equal to the applied magnetic field $H$. The
latter is exactly solvable \cite{SGM1,SGM2} and exhibits two distict
phases. If $H$ is less than the critical value $H_{c,1}$ the model
remains gapped, while it becomes critical for $H>H_{c,1}$. 
On the basis of these observations we expect altogether four different
phases. The interaction term ${\cal H}_{\rm int}$ affects the value of
the critical field $H_{c,1}$ and renormalizes the parameters in
${\cal H}_3$ and ${\cal H}_B$. In addition to the terms written in
\fr{HFT2} further (marginal or irrelevant) interactions will be
generated by integrating out high energy degrees of freedom in the
underlying lattice model. In particular, the marginal interaction
${\cal H}_{\rm marginal}=\partial_x\Phi R_3L_3$ appears to be
compatible with all symmetries and therefore ought to be
generated. While this term is unimportant as long as the Ising model
described by ${\cal   H}_3$ remains off-critical, it could modify the
Ising transition itself. The effects of this interaction term have
recently been analyzed in the context of a band-filling transition in
a two-subband quantum wire by Sitte et al. \cite{Sitte} Based on a 1-loop
renormalization group calculation it was suggested that 
${\cal H}_{\rm marginal}$ alters the nature of the quantum phase
transition and leads to a scaling behavior that differs from that of 
${\cal H}_3+{\cal H}_B$ for very large length scales. Due to the
non-scalar nature of ${\cal H}_{\rm marginal}$ these length scales are
expected to be much larger than the system sizes used in our DMRG
computations. In the following we therefore will neglect
${\cal H}_{\rm marginal}$.

\subsection{Weak magnetic fields $H<H_{c,1}(\theta)$}
\label{sec:fieldtheoryweakfields}
For weak magnetic fields $H<H_{c,1}(\theta)$ the model \fr{HFT2} is fully
gapped. The system remains unmagnetized. There are two phases, which
are distinguished by the sign of the mass term in ${\cal
  H}_3$.\cite{amt,shura} For $\theta-\theta_{TB}>0$ we have $m>0$ and
the Ising model described by ${\cal H}_3$ is in its disordered
phase. This corresponds to the Haldane phase. For
$\theta-\theta_{TB}<0$ we have $m<0$ and the Ising model described by
${\cal H}_3$ is in its ordered phase. This corresponds to the
dimerized phase.

\subsection{High fields $H>H_{c,1}(\theta)$}
\label{sec:fieldtheoryhighfields}
When the magnetic field exceeds the critical value $H_{c,1}(\theta)$
the bosonic degrees of freedom described by ${\cal H}_B$ enter a
gapless Luttinger liquid phase. For $H>H_{c,1}(\theta)$ we end up with an
effective Hamiltonian of the form
\bea
{\cal H}&=&\frac{i\tilde{v}}{2}\left[L_3\partial_xL_3-R_3\partial_xR_3\right]
-i\tilde{m}\ R_3L_3\nn
&+&\frac{\tilde{v}'}{2}\left[(\partial_x\tilde{\Phi})^2
+(\partial_x\tilde{\Theta})^2\right] ,
\label{HFT3}
\eea
where the parameters $\tilde{v}$, $\tilde{v}'$ and $\tilde{m}$ depend
on the magnetic field $H$ and $\theta$.
This can be seen as follows. We may remove the magnetic field term in
\fr{HFT2} by the field redefinitions
$\Phi'=\Phi+\frac{K}{v'\sqrt{\pi}}Hx$, $\Theta'=\Theta$.
In terms of the new fields the cosine term in ${\cal H}_B$
is oscillating in $x$ and for sufficiently large $H$ (compared to $m$)
drops out of the Hamiltonian $\int dx\ {\cal H}$ at low
energies. Carrying out a unitary rescaling of $\Phi'$ and $\Theta'$
then leads to \fr{HFT3} (where we have also taken into account the
renormalization of the various parameters in a strong field).
The lattice spin operators are expressed in terms of the new fields as
\bea
S^\pm_j&\sim&(-1)^j\exp\Big({\pm
  i\frac{\pi}{\beta}\tilde{\Theta}}\Big)\ \mu^3+\ldots\ ,\nn
S^z&\sim&\cos(\beta\tilde{\Phi}+\pi (1-M)x)\ \sigma^3+\ldots\ ,\nn
(S^\pm_j)^2&\sim&\exp\Big({\pm
  i\frac{2\pi}{\beta}\tilde{\Theta}}\Big)+\ldots\ ,
\label{spinops}
\eea
where $\beta>\beta(H_{c,1})=\sqrt{\pi}$ depends on
the applied magnetic field $H$ \cite{haldane1983PRL,Konik2002} in a way that
cannot be easily calculated from within the field theory framework.
In \fr{spinops} we have only written the contributions with the
smallest scaling dimensions at the TB point.

The form of \fr{HFT3} shows that there are two phases separated by an
Ising phase transition, which occurs when $\tilde{m}$ is tuned to
zero. We are now in a position to describe the behaviour of
correlation functions in these phases.

\subsubsection{Haldane Phase in a field}
\label{sec:Haldanephasefieldtheory}
The case $\tilde{m}>0$ corresponds to the Haldane phase in a (strong)
magnetic field. We note that the closely related case $\theta=0$,
i.e. the spin-1 Heisenberg model in a field, has been discussed
previously by several authors, see
e.g. Refs [\onlinecite{Affleck1991,amt,Konik2002,FZ}]. The Ising model
described by $L_3$, $R_3$ is in its disordered phase so that
\be
\langle\sigma^3\rangle=0\ ,\qquad
\langle\mu^3\rangle\neq0\ .
\ee
More precisely, the expectation value of the Ising disorder operator
scales as 
\be
\langle\mu^3\rangle\propto |\tilde{m}|^\frac{1}{8}.
\ee
If we consider the Ising mass as a function of the parameter $\theta$
for fixed magnetic field $H$, we have very close to the critical point
$\theta_c(H)$
\be
\tilde{m}\propto |\theta-\theta_c|.
\ee
Using the expression given in \fr{spinops} gives an exponentially
decaying contribution to the zz spin correlations. The leading
long-distance behaviour is therefore due to the terms
$\partial_x\tilde{\Phi}$ and $\sin(2\beta\tilde{\Phi}+2\pi Mx)$
\cite{Konik2002,FZ}, which gives
\be
C_{S}^{\rm long}(i,j)\sim\frac{A}{(i-j)^2}+
\frac{B\cos\big(2\pi M(i-j)\big)}{(i-j)^{2\beta^2/\pi}}.
\ee
Here the oscillatory contribution is always subleading as
$\beta>\sqrt{\pi}$.
The dominant correlations are the transverse spin correlations
\bea
C_{S}^{\rm trans}(i,j)&\sim&(-1)^{i-j}
\langle \exp\big(i\frac{\pi}{\beta}\tilde{\Theta}(x)\big)
\exp\big(-i\frac{\pi}{\beta}\tilde{\Theta}(0)\big)\rangle\nn
&\sim&(-1)^{i-j}\ (i-j)^{-\pi/(2\beta^2)}.
\label{eq:FT_cstrans1}
\eea
Hence we expect a correlation exponent for transverse spin correlations
\be
\frac{\pi}{2\beta^2}<\frac{1}{2}.
\ee
The high-field phase is an attractive Luttinger liquid with dominant
transverse spin correlations, in agreement with Ref.~\onlinecite{Konik2002}. The
long-distance asymptotics of the quadrupolar correlations is
\bea
C_{Q,2}(i,j)&\sim&
\langle \exp\big(i\frac{2\pi}{\beta}\tilde{\Theta}(x)\big)
\exp\big(-i\frac{2\pi}{\beta}\tilde{\Theta}(0)\big)\rangle\nn
&\sim&(i-j)^{-2\pi/\beta^2}.
\eea
Hence we have
\be
\lim_{|i-j|\to\infty}\frac{|C^{\rm trans}_S(i,j)|^4}{C_{Q,2}(i,j)}
={\rm const}\propto
|\theta-\theta_c|,
\ee
in agreement with the numerical results shown in
Fig.~\ref{fig:Isingorderparameter}(a). 

In order to make contact with DMRG calculations it is useful to
consider Friedel oscillations in $S^z_j$ for a system with
boundaries. At the Heisenberg point $\theta=0$ we may derive the
Luttinger liquid description of the magnetized phase of the open chain
using a strong coupling analysis as in
Refs~\onlinecite{FZ,Konik2002}. This results in an effective spin-1/2
Heisenberg XXZ chain with equal boundary magnetic fields on both
ends. Standard bosonization methods then give
\be
\langle S^z_j\rangle\sim M+A\frac{\sin\Big(2\pi \tilde{M}j+\pi\varphi\Big)}{\left|\frac{N+1}{\pi}\sin\big(\frac{\pi
    j}{N+1}\big)\right|^{\beta^2/\pi}},
\label{friedelhaldane}
\ee
where $N$ is the length of the chain and
\be
\tilde{M}=M+\frac{\frac{1}{2}-M-\varphi}{N+1}.
\ee
 We expect the form of
\fr{friedelhaldane} to hold also away from $\theta=0$.

\subsubsection{Dimerized Phase in a field}
\label{sec:fieldtheorydimerizedphase}
Here we have $\tilde{m}<0$ and the Ising model described by $L_3$, $R_3$ is in
its ordered phase. Hence we have
\be
\langle\sigma^3\rangle\neq0\ ,\qquad
\langle\mu^3\rangle=0\ .
\ee
As a result the contribution to the transverse spin correlations
due to the ``leading'' operator identified in \fr{spinops} decay
exponentially, as do other contributions we have considered.
The asymptotics of the zz spin correlator is
\bea
C^{\rm long}_S(i,j)&\sim& (i-j)^{-\beta^2/(2\pi)}\cos\big(\pi(1-M)(i-j)\big)\nn
&\equiv&(i-j)^{-\Delta_1}\cos\big(\pi(1-M)(i-j)\big) .
\eea
On the other hand, the quadrupolar correlations behave as
\bea
C_{Q,2}(i,j)&\sim&(i-j)^{-2\pi/\beta^2}\equiv(i-j)^{-\Delta_2}\ .
\eea
Just above $H_{c,1}$ we have $\beta\approx\sqrt{\pi}$, which
implies that
\be
\Delta_2\approx2\ ,\qquad \Delta_1\approx\frac{1}{2}.
\ee
These agree with the weak-field limit of the Bethe-ansatz analysis
above. We furthermore know from the Bethe-ansatz analysis that with
increasing field the parameter $\beta$ grows, which implies that
$\Delta_2$ decreases and $\Delta_1$ increases. For sufficiently strong
fields the quadrupolar correlations become dominant.

For an open chain we expect Friedel oscillations of the form
\be
\langle S^z_j\rangle\sim
M+C\frac{\sin\Big(\pi(1-\tilde{M})j+\pi\varphi'\Big)}{\left|\frac{N+1}{\pi}\sin\big(\frac{\pi j}{N+1}\big)\right|^{\beta^2/4\pi}},
\label{friedeldimer}
\ee
where for $M=m/N$ with even $m,N$ we find
\be
\tilde{M}=M+\frac{2\varphi'-M}{N+1}.
\ee
\subsubsection{Ising Transition}
\label{sec:fieldtheorytransition}
At the transition we have $\tilde{m}=0$ and the Ising model is
critical. As a result the spin correlators acquire additional
power-law factors and become
\bea
C_{s}^{\rm trans}(i,j)&\sim&
(-1)^{i-j}\ (i-j)^{-\frac{1}{4}-\pi/(2\beta^2)}\ ,  \label{eq:FT_cstrans2} \\
C_{s}^{\rm long}(i,j)&\sim&
(i-j)^{-\frac{1}{4}-\beta^2/(2\pi)}
\cos\big(\pi(1-M)(i-j)\big),\nn
C_{Q,2}(i,j)&\sim&(i-j)^{-2\pi/\beta^2}\ . \nonumber
\eea

Note that this implies that $R(x)$ [Eq.~\ref{eq:ratio}] decays $\sim 1/x$, while comparing expressions (\ref{eq:FT_cstrans1}) and (\ref{eq:FT_cstrans2}) shows that the exponent of the transverse spin correlations jumps upon entering the magnetized Haldane phase by a value of $1/4$ independent of the value of the magnetization. In contrast, the exponent of $C_{Q,2}$ changes continuously across the transition. These findings are all in agreement with the numerical results presented in Figs.~\ref{fig:exponents}(b), (c) and \ref{fig:Isingorderparameter}.

In conclusion, our combined numerical and field theoretical analysis supports the picture that the Ising transition identified in the fermionic $S=3/2$ attractive Hubbard model finds a corresponding counterpart in the pair-unbinding transition of the $S=1$ BLBQ chain at finite magnetic fields.

\section{Summary and Conclusion}
\label{sec:summary}
To summarize, by combining extensive DMRG calculations, Bethe ansatz and field theoretical arguments we have determined the complete phase diagram of the $S=1$ BLBQ Heisenberg chain in a magnetic field.
At finite magnetizations, it consists of five phases, three single component LL phases and two two-component LL phases.
Two of the single component LL phases appear when polarizing the system starting from the dimerized phase at negative biquadratic
interactions.
At large enough fields, the LL realized in this parameter region is a ferroquadrupolar LL, which is connected to a SDW-type of LL at lower fields via a crossover line. In the whole region, the gap to single magnon excitations is finite, and both LL phases are characterized by a quasi-condensate of bound pairs of magnons. These two single-channel LLs of pairs of magnons are connected by a continuous transition to the more standard single component LL phase of single magnons that appears when polarizing the Haldane phase. We determined the transition to belong to the Ising universality class with a central charge of  3/2 due to the contribution of the adjacent LL phases. This transition emerges at the TB point at zero field, showing that the magnetic field moves the universality class from SU(2)$_2$ WZWN at zero field to Luttinger liquid plus Ising at finite fields.
The two-component LL phases show up for large positive biquadratic interaction (and positive bilinear interaction). They are separated by
a magnetization kink from the magnetized Haldane phase and are characterized by dominant incommensurate correlations of
transverse magnetic resp. quadrupolar type. It is remarkable that they reflect to a certain extend the behavior at zero field. In particular, the spin-nematic character of the LL phase identified at zero field in the region $\pi/4 \leq \theta \leq \pi/2$ survives when applying a magnetic field.
It is our hope that this rich phase diagram will further
motivate the search for experimental realizations of this model both, in quantum magnetic materials as well as in systems of ultracold atomic gases on optical lattices. 

\section*{Acknowledgements}
We acknowledge useful discussions with E.~Boulat, S.~Capponi, K.~Penc, J.-D.~Picon, J.~Sudan, and T.A.~T\'oth. 
This work has been supported by the Swiss National Fund and by MaNEP. FHLE thanks the MPIPKS Dresden for its hospitality and acknowledges funding by EPSRC grant EP/D050952/1. SRM acknowledges financial support by PIF-NSF (grant No. 0904017). This research was supported in part by the National Science Foundation under Grant No. NSF PHY05-51164.  
\appendix
\section{Bethe Ansatz Analysis at the Takhtajan-Babujian Point with a Magnetic Field}
\label{appendix}

In this Appendix we describe in more detail the Bethe ansatz analysis
of the $S=1$ BLBQ Heisenberg chain in a magnetic field at the TB point
$\theta_{TB}=-\frac{\pi}{4}$. At this point the Hamiltonian takes the
form
\be
H=\frac{J}{\sqrt{2}}\sum_{j=1}^L{\bf S}_j\cdot{\bf S}_{j+1}
-\left({\bf S}_j\cdot{\bf S}_{j+1}\right)^2
-HS^z,
\label{H}
\ee
and is known to be solvable by Bethe ansatz \cite{takhtajan,babujian_nuclphysB,TsvelikNucPhysB} for arbitrary
values of the magnetic field $H$. The ground state is described in
terms of the integral equation
\bea
\rho_2(\lambda)&=&a_1(\lambda)+a_3(\lambda)\nn
&-&\int_{-A}^Ad\mu\left[2a_2(\lambda-\mu)
+a_4(\lambda-\mu)\right]\rho_2(\mu),
\eea
where
\be
a_n(\lambda)=\frac{1}{2\pi}\ \frac{2n}{n^2+\lambda^2}.
\label{eq:anlambda}
\ee
The integration boundary $A$ is fixed by the condition
\be
\eps_2(A)=0\ ,
\ee
where the dressed energy $\epsilon_2(\lambda)$ is a solution of the
integral equation
\be
\eps_2(\lambda)=\eps_2^{(0)}(\lambda)-\int_{-A}^Ad\mu\left[2a_2(\lambda-\mu)
+a_4(\lambda-\mu)\right]\eps_2(\mu).
\ee
Here the ``bare energy'' is given by
\be
\eps^{(0)}_2(\lambda)=-\frac{8\pi J}{\sqrt{2}}[a_1(\lambda)+a_3(\lambda)]+2H.
\ee
Ground state energy and magnetization per site are
\bea
M&=&1-2\int_{-A}^Ad\lambda\ \rho_2(\lambda)\ ,\nn
e&=&\int_{-A}^Ad\lambda\ \eps^{(0)}_2(\lambda)\
\rho_2(\lambda)\ .
\eea
For zero field the determination of the finite-size spectrum of
low-lying excitations is difficult because the ground state is made
from complex solutions of the Bethe ansatz equations (``2-strings'').
\cite{strings1,strings2,strings3,strings4} In a finite field matters are simpler and following
the standard analysis \cite{vladbook} we can establish that the
finite-size spectrum of low-lying excited states is given by
\bea
\Delta E&=&\frac{2\pi v}{L}\left[\frac{(\Delta
    N)^2}{4Z^2}+(Zd)^2+N^++N^-\right]\ ,\\
\Delta P&=&2k_F d+\frac{2\pi}{L}\left[N^+-N^-+d\Delta N\right].
\label{gaussian}
\eea
Here $\Delta N$ and $d$ are integers and the dressed charge $Z=\xi(A)$
is calculated from the integral equation
\be
\xi(\lambda)=1-\int_{-A}^Ad\mu\left[2a_2(\lambda-\mu)+a_4(\lambda-\mu)\right]
\xi(\mu).
\ee
\begin{figure}[b]
\includegraphics[width=0.45\textwidth]{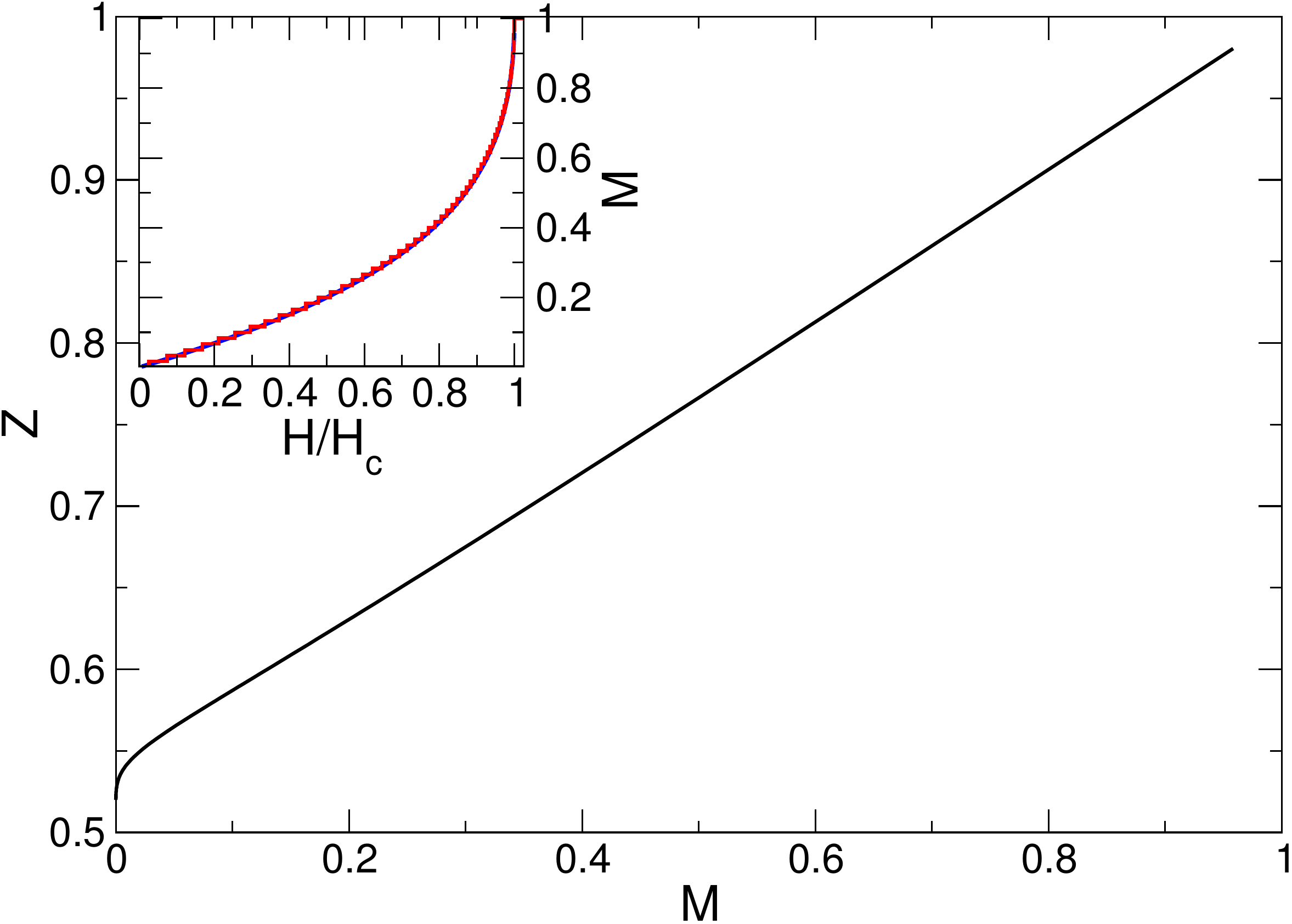}
\caption{(Color online) Dressed charge $Z$ as a function of the magnetization per
  site $M$ as obtained by Bethe ansatz. Inset: magnetization per site
  $M$ as a function of the applied magnetic field (in units of the
  saturation field $H_c$) as obtained by Bethe ansatz (blue straight line) and the DMRG (red steps).}
\label{fig:Z}
\end{figure}
The result is shown in Fig.~\ref{fig:Z}, which depicts the dressed
charge $Z$ as a function of the magnetization $M$.
The integer $\Delta N$ is related to the z-component of the spin by
\be
\delta S^z=-2\Delta N.
\ee
The ``Fermi momentum'' $k_F$ is related to the magnetization per site
by
\be
k_F=\frac{\pi(1-M)}{2}.
\ee
The spectrum \fr{gaussian} describes a Gaussian model, which implies that
the asymptotic behaviour of correlation functions takes the form
\cite{vladbook}
\bea
\langle{\cal O}(t,x) {\cal O}^\dagger(0,0)\rangle&=&\sum_{d,\Delta N,N^\pm}
C(d,\Delta N,N^\pm)\ e^{-2ixk_Fd}\nn
&\times&\! (x-ivt)^{-2\Delta^+}(x+ivt)^{-2\Delta^-}\!,
\eea
where
\bea
2\Delta^\pm=2N^\pm+\left(\frac{\Delta N}{2Z}\pm Zd\right)^2.
\eea
Which of the amplitudes $C(d,\Delta N,N^\pm)$ are non-zero depends on
the operator under consideration.
The smallest correlation exponents for scalar operators are obtained
by the choices
\bea
&&\Delta N=0,\ d=1 \longrightarrow \Delta_1=2Z^2\ ,\nn
&&d=0,\ \Delta N=1 \longrightarrow \Delta_2=\frac{1}{2Z^2}.
\eea
The z-component of the spin $S^z_j$ is sensitive only to states with
$\Delta N=0$, while $(S_j^-)^2$ changes the total spin by 2 and hence
couples to states with $\Delta N=1$.
This analysis then suggests that the leading long-distance behaviour
of correlation functions is of the form
\bea
C_S^{\rm long}(i,j)&\sim& (i-j)^{-\Delta_1}\cos\big(\pi(1-M)(i-j)\big),\nn
C_{Q,2}(i,j)&\sim& (i-j)^{-\Delta_2}\ ,
\eea
The exponents $\Delta_{1,2}$ are shown as functions of the applied
magnetic field in Fig.~\ref{fig:Delta_M}. We see that for weak fields the
longitudinal spin correlations dominate, while for strong fields the
quadrupolar correlations decay more slowly. The crossover occurs at a
magnetization per site of $M_c\approx 0.37$.

\section{Bethe Ansatz solution for $\theta=\theta_{ULS}=\frac{\pi}{4}$}
\label{appendix2}
Here we describe in more detail the Bethe Ansatz analysis
of the SU(3) Uimin-Lai-Sutherland model \cite{uimin,lai,sutherland}
in a magnetic field, which corresponds to the point
$\theta_{ULS}=\frac{\pi}{4}$. For this value of $\theta$ the
Hamiltonian takes the form  
\be
H=\frac{J}{\sqrt{2}}\sum_{j=1}^L{\bf S}_j\cdot{\bf S}_{j+1}
+\left({\bf S}_j\cdot{\bf S}_{j+1}\right)^2
-hS^z,
\label{Hsu3}
\ee
and is known to be solvable by Bethe ansatz for arbitrary
values of the magnetic field $H$. The critical properties of the model
have been previously analyzed in Ref.~[\onlinecite{fath_littlewood}] and we begin
by summarizing the results obtained there. 

\subsection{Ground State Properties}
\begin{figure}[ht]
\includegraphics[width=0.48\textwidth]{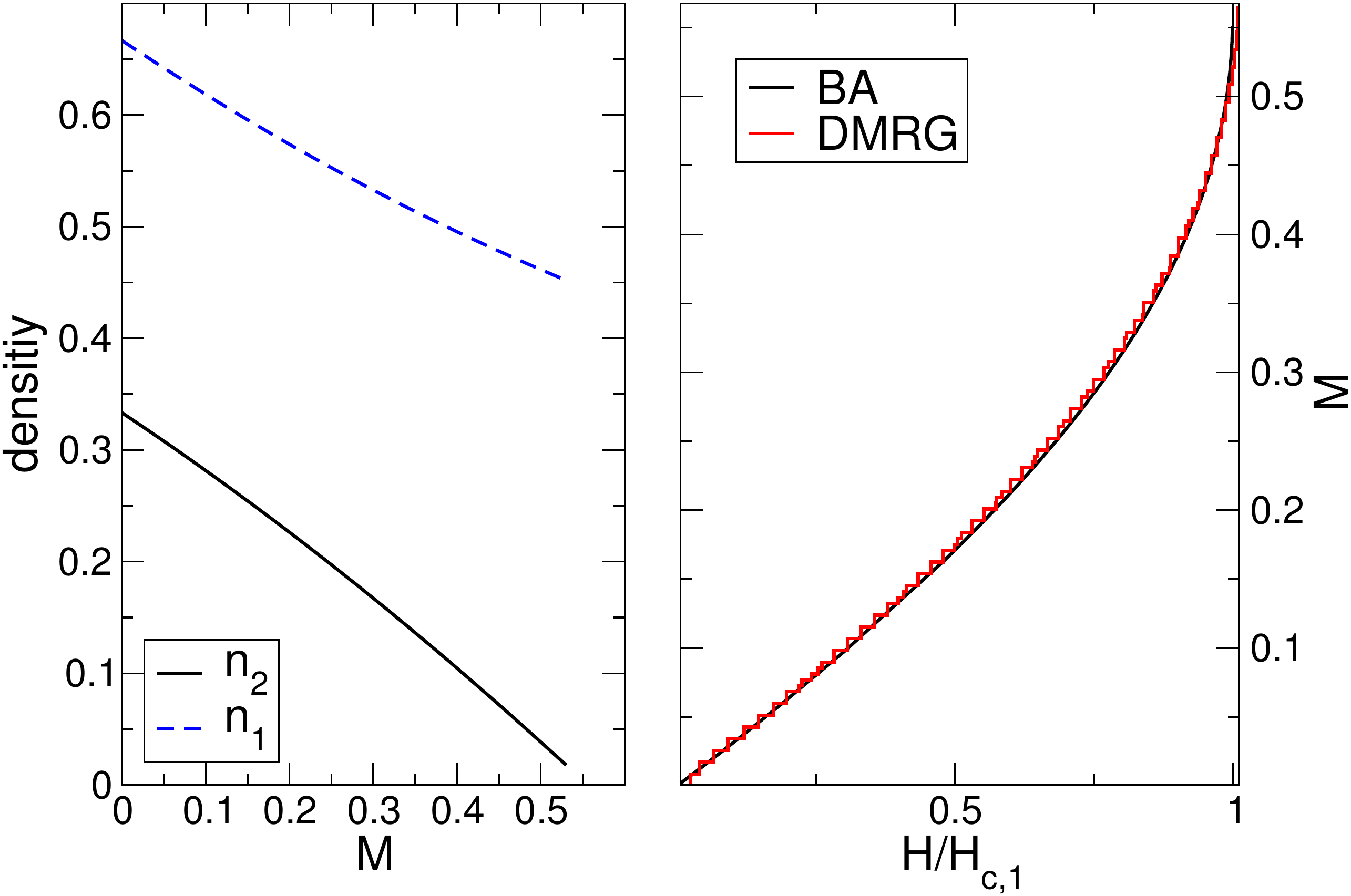}
\caption{(Color online) 
Densities $n_{1,2}$ as functions of the magnetization per site as obtained by Bethe ansatz and magnetization per site as a function of the applied magnetic fieldÊ
as obtained by Bethe ansatz (black straight line) and DMRG (red steps). 
} 
\label{fig:su3dens}
\end{figure}

The ground state is described in
terms of the coupled integral equations
\bea
\rho_a(\lambda)&=&\rho_a^{(0)}(\lambda)+\sum_{b=1}^2\int_{-A_b}^{A_b}d\mu
K_{ab}(\lambda-\mu)\rho_b(\mu)\ ,
\label{dressedden}
\eea
where 
\bea
K_{12}(\lambda)&=&K_{21}(\lambda)=a_1(\lambda)\ ,\\
K_{11}(\lambda)&=&K_{22}(\lambda)=-a_2(\lambda)\ ,\\
\rho_a^{(0)}(\lambda)&=&-2\pi\delta_{a1}a_1(\lambda), 
\eea
with $a_n(\lambda)$ defined in Eq.~(\ref{eq:anlambda}).  
By virtue of the enhanced $SU(3)$ symmetry of the model
(\ref{Hsu3}) in zero field the numbers $M_{\sigma}$ of
$\sigma$ spins ($\sigma=1,0,-1$) are good quantum numbers. The
z-component of total spin is one of the Cartan generators of SU(3) and
hence $M_\sigma$ remain good quantum numbers even in the presence of a
magnetic field. By definition we 
have $L=M_1+M_0+M_{-1}$ and in the ground state we have
\bea
n_1&=&\frac{N_1}{L}=\frac{M_0+M_{-1}}{L}=\int_{-A_1}^{A_1}d\lambda \rho_1(\lambda)\ ,\\
n_2&=&\frac{N_2}{L}=\frac{M_{-1}}{L}=\int_{-A_2}^{A_2}d\lambda \rho_2(\lambda)\ .
\label{n12}
\eea
The conditions \fr{n12} fix the integration boundaries $A_{1,2}$ as
functions of the densities $n_{1,2}$. The magnetization per site is
\bea
M&=&\frac{M_1-M_{-1}}{L}=1-\sum_{b=1}^2\int_{-A_b}^{A_b}d\lambda\
\rho_b(\lambda)\nn
&=&1-n_1-n_2.
\eea
The integration boundaries $A_{1,2}$ are determined by the applied
magnetic field through the conditions
\be
\eps_b(A_b)=0\ ,
\ee
where the dressed energies $\epsilon_b(\lambda)$ are solutions of the
coupled integral equations
\bea
\eps_a(\lambda)&=&\eps_a^{(0)}(\lambda)+\sum_{b=1}^2\int_{-A_b}^{A_b}d\mu
K_{ab}(\lambda-\mu)\eps_b(\mu)\ .
\label{dresseden}
\eea
Here the ``bare energies'' are given by
\bea
\eps^{(0)}_1(\lambda)&=&-2\pi a_1(\lambda)+h\ ,\\
\eps^{(0)}_2(\lambda)&=&h\ .
\eea
As a function of magnetic field there are four distinct regimes:
\begin{enumerate}
\item{} $h=0$: as a result of the enhanced symmetry the low-energy
  physics is described by the $SU_1(3)$ WZNW model. The central charge
  is $c=2$.
\item{} $0<h<h_{c,1}$: the model remains in a quantum critical
  phase. Universal properties are described by a two-component
  Luttinger liquid. The central charge is $c=2$, but the symmetry is
  reduced as compared to $h=0$. When $h$ approaches $h_{c,1}$ the
  cutoff of one of the Luttinger liquids goes to zero.
\item{} $h_{c,1}<h<h_{c,2}$: the low energy physics is described by a
  $c=1$ one-component Luttinger liquid.
\item{} $h_{c,2}<h$: the ground state is fully polarized and all
  excitations have a gap.
\end{enumerate}
In the following we concentrate on the two-component Luttinger liquid
regime $0<h<h_{c,1}$. In Fig.\ref{fig:su3dens} we plot the
magnetization per site as a function of the applied magnetic field and
the densities $n_{1,2}$ as functions of the magnetization. We see that
at $h_{c,1}$ the density of $S^z=-1$ spins becomes zero.

\subsection{Low-lying excitations for $0<h<h_{c,1}$}
As we are dealing with a quantum critical theory there are gapless
excitations. In a finite volume $L$ the spectrum of low-lying excited
states scales as $L^{-1}$ and is related to the operator content of
the underlying conformal field theory \cite{Cardy86}.
The finite-size energies and momenta of low-lying excitations can be
determined by standard methods \cite{FSspectrum1,FSspectrum2,book_hubbardmodel} from the Bethe
Ansatz solution with the result \cite{fath_littlewood}
\begin{equation}
\label{gftem}
\begin{aligned}
   E(\Delta {\bf N},{\bf d}) - E_0 &= {{2 \pi} \over L} \sum_{a=1}^2
v_a (\Delta_a^+ + \Delta_a^-)        + o({1\over L})\ ,\\
   P(\Delta {\bf N},{\bf d}) - P_0 &= {{2 \pi} \over L} \sum_{a=1}^2
        \Delta_a^+ - \Delta_a^-\\
	&+2\pi (n_1d_1+n_2 d_2)+\pi \Delta N_1,
\end{aligned}
\end{equation}
where the conformal dimensions $\Delta_{1,2}^\pm$ are expressed as
\begin{widetext}
\begin{equation}\label{cf:gdim}
\begin{aligned}
   \Delta_1^\pm(\Delta{\bf N}, {\bf d},N_1^\pm) &= \frac{1}{2}\left(
       Z_{11} d_1 + Z_{21} d_2 \pm {{Z_{22} \Delta N_1 - Z_{12}
       \Delta N_2}\over{2 \det Z}} \right)^2 + N_1^\pm,
   \\
   \Delta_2^\pm(\Delta{\bf N}, {\bf d},N_2^\pm) &= \frac{1}{2}\left(
       Z_{12} d_1 + Z_{22} d_2 \pm {{Z_{11} \Delta N_2 - Z_{21}
       \Delta N_1}\over{2 \det Z}} \right)^2 + N_2^\pm.
\end{aligned}
\end{equation}
\end{widetext}
Here $N_a^\pm$ and $\Delta N_{1,2}$ are integer numbers,
\be
d_1=\frac{\Delta N_{2}}{2}\ {\rm mod}\ 1,\qquad d_2=\frac{\Delta
  N_{1}}{2}\ {\rm mod}\ 1,
\ee
and $v_{1,2}$ are Fermi velocities of the
two types of elementary excitations. They are given in terms of the
integral equations \fr{dresseden}, \fr{dressedden} by 
\be
v_a=\frac{\eps'_a(A_a)}{2\pi\rho_a(A_a)},
\ee
where $\eps'_a(\lambda)$ are the derivatives of the dressed energies.
Finally, $Z_{ab}$ are the elements of the \emph{dressed charge matrix}
\be
Z=
\begin{pmatrix}
\xi_{11}(A_1) & \xi_{12}(A_2)\cr
\xi_{21}(A_1) & \xi_{22}(A_2)
\end{pmatrix}\ ,
\label{Z}
\ee
where $\xi_{ab}$ fulfil the
set of coupled integral equations 
\be
\xi_{ab}(\lambda)=\delta_{ab}+\sum_{c=1}^2\int_{-A_c}^{A_c}d\mu\
\xi_{ac}(\mu)\ K_{cb}(\mu-\lambda)\ .
\label{xi}
\ee

\begin{figure}[ht]
\includegraphics[width=0.48\textwidth]{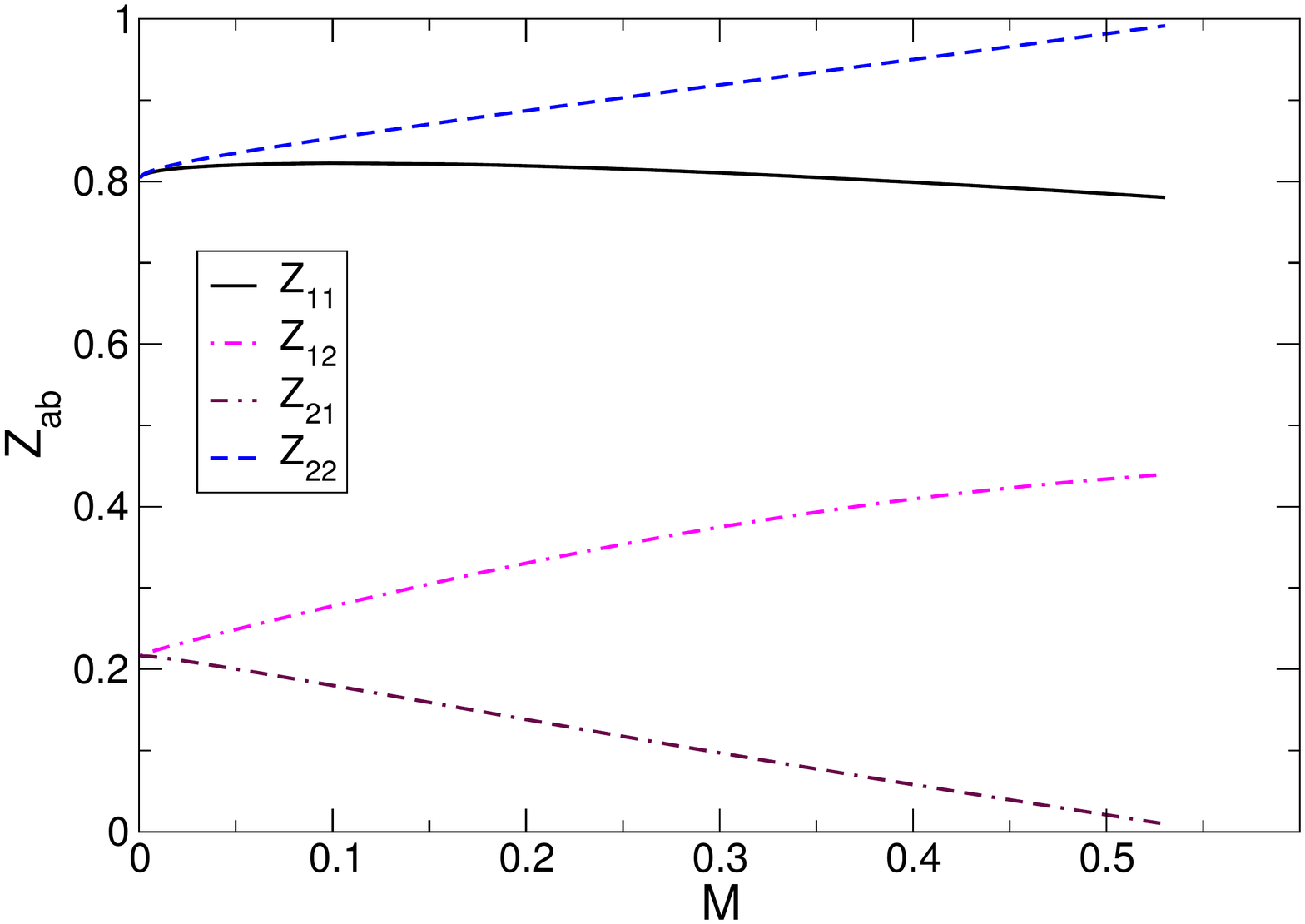}
\caption{(Color online) Elements of the dressed charge matrix $Z$ as functions of the
  magnetization per site $M$ as obtained by Bethe Ansatz. }
\label{fig:Zab}
\end{figure}
In Fig.\ref{fig:Zab} we plot the elements of the dressed charge matrix
$Z$ as functions of the magnetization. 

\subsection{Long-distance asymptotics of correlation functions}
As the critical behaviour is described by a two-component Luttinger
liquid the asymptotic behaviour of correlation functions can be
extracted from the finite-size spectrum following the analysis of
Frahm and Korepin for the Hubbard model\cite{FrKo91,book_hubbardmodel}. The
asymptotic behaviour of the two-point function of a local operator
${\cal O}$ is given by
\bea
\langle{\cal O}(x) {\cal O}^\dagger(0)\rangle&=&\sum_{{\bf d},\Delta
  {\bf N},{\bf N}^\pm}
C({\bf d},\Delta {\bf N},{\bf N}^\pm)\ 
x^{-\Delta}\nn
&&\times\ e^{-2\pi i (n_1d_1+n_2 d_2+\frac{1}{2}\Delta N_1)},
\label{asymptotics}
\eea
where the exponents $\Delta$ are related to the finite size energies by
\bea
\Delta({\bf d},{\bf \Delta N},{\bf N}^+,{\bf N}^-)
=2\Delta_1^++2\Delta_1^-+2\Delta_2^++2\Delta_2^-.
\eea
For a given operator ${\cal O}$ certain amplitudes $C(d,\Delta
N,N^\pm)$ will be zero due to continuous or discrete symmetries, which
sometimes are not entirely obvious\cite{EF99}.

For later use we define a number of momenta characterizing the
oscillatory behavior of correlation functions
\bea
P_1&=&2\pi(n_1-n_2)\ ,\nn
P_2&=&\pi(1-n_2)\ ,\nn
P_3&=&\pi(1+n_1-n_2).
\eea

\subsection{Longitudinal Spin Correlations}
As $S^z_j$ does not change the total $z$-component of spin only
intermediate states for which $S^z=L-N_1-N_2$ is the same as in the
ground state will contribute to the correlation function. Hence the
longitudinal correlations are characterized by quantum numbers subject
to the selection rule 
\be
\Delta N_1+\Delta N_2=0.
\ee
The smallest exponents are then obtained by the choices
(1) $\Delta N_{1,2}=0$, $d_1=\pm 1$, $d_2=0$, $N^\pm=0$,\\
(2) $\Delta N_{1,2}=0$, $d_1=0$, $d_2=\pm 1$, $N^\pm=0$,\\
(3) $\Delta N_{1,2}=0$, $d_1=-d_2=\pm 1$, $N^\pm=0$,\\
(4) $\Delta N_{1,2}=0$, $d_1=d_2=0$, $N^-=0$, $N^+=1$,\\
(5) $\Delta N_{1,2}=0$, $d_1=d_2=0$, $N^-=1$, $N^+=0$.

This leads to the following form for $C_S^{\rm long}(i,j)$
\bea
C_S^{\rm long}(i,j)&\sim& C_1
(i-j)^{-\Delta^{\rm long}_1}\cos\big(2\pi n_1(i-j)\big)\nn
&+&C_2(i-j)^{-\Delta^{\rm long}_2}\cos\big(2\pi n_2(i-j)\big)\nn
&+&C_3(i-j)^{-\Delta^{\rm long}_3}\cos\big(P_1(i-j)\big)\nn
&+&C_4(i-j)^{-2}+\ldots,
\eea
where
\bea
\Delta^{\rm long}_1&=&2(Z_{11}^2+Z_{12}^2)\ ,\nn
\Delta^{\rm long}_2&=&2(Z_{21}^2+Z_{22}^2)\ ,\nn
\Delta^{\rm long}_3&=&2(Z_{11}-Z_{21})^2+2(Z_{12}-Z_{22})^2.
\eea
The magnetization dependence of $\Delta^{\rm long}_{1,2,3}$ is shown
in Fig.~\ref{fig:su3exp}.
\subsection{Transverse Spin Correlations}
In the transverse spin correlator only intermediate states with 
\be
\Delta N_1+\Delta N_2=\pm 1
\ee
contribute. The smallest exponents are then obtained by the choices 

\noindent
(1) $\Delta N_1=\pm 1$, $d_2=\pm\frac{1}{2}$, 
$\Delta N_2=d_1=N^\pm=0$.\\
(2) $\Delta N_2=\pm 1$, $d_1=\pm\frac{1}{2}$, 
$\Delta N_1=d_2=N^\pm=0$.\\
This leads to the following form for $C_S^{\rm trans}(i,j)$
\bea
C_S^{\rm trans}(i,j)&\sim& 
D_1(i-j)^{-\Delta^{\rm trans}_1}\cos\big(P_2(i-j)\big)\nn
&+&D_2(i-j)^{-\Delta^{\rm trans}_2}\cos(\pi n_1(i-j)\nn
&+&\ldots,
\eea
where the exponents are given by
\bea
\Delta^{\rm trans}_1&=&[Z_{21}^2+Z_{22}^2]\frac{1+\det^2Z}{2\det^2Z}\ ,\nn
\Delta^{\rm trans}_2&=&[Z_{12}^2+Z_{11}^2]\frac{1+\det^2Z}{2\det^2Z}\ .
\eea
The magnetization dependence of $\Delta^{\rm trans}_{1,2}$ is shown
in Fig.~\ref{fig:su3exp}. We see that the two exponents are comparable
in magnitude but $\Delta_2^{\rm  trans}<\Delta_1^{\rm trans}$. 
\subsection{Quadrupolar Correlations}
Here the operator ${\cal O}$ in (\ref{asymptotics}) changes the
z-component of total spin by $\pm 2$, so that we  need to consider
intermediate states with $\Delta N_1+\Delta N_2=\pm 2$. 
The smallest exponents are then obtained by the choices 

\noindent
(1) $\Delta N_1=\Delta N_2=\pm 1$, $d_1=-d_2=\pm\frac{1}{2}$, $N^\pm=0$,\\
(2) $\Delta N_2=2$, $\Delta N_1=d_{1,2}=N^\pm=0$,\\

\begin{figure}[ht]
\includegraphics[width=0.48\textwidth]{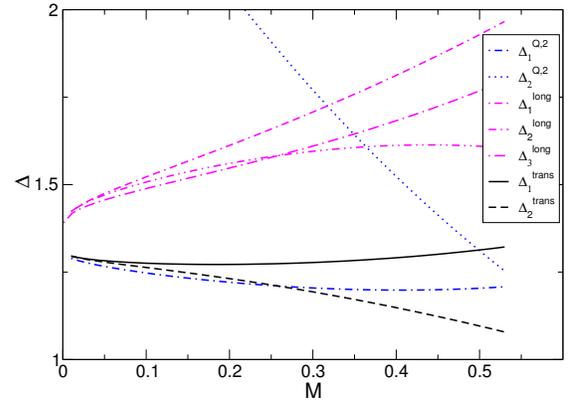}
\caption{(Color online) Exponents characterizing the power-law decays of the various
  correlators as functions of the magnetization per site.}
\label{fig:su3exp}
\end{figure}

This leads to the following form for $C_Q(i,j)$
\bea
C_{Q,2}(i,j)&\sim& 
E_1(i-j)^{-\Delta^{(1)}_{Q,2}}\cos\big(P_3(i-j)\big)\nn
&+&
E_2(i-j)^{-\Delta^{(2)}_{Q,2}}\nn
+\ldots,
\eea
where the exponents are given by
\bea
\Delta^{Q,2}_1&=&[(Z_{11}-Z_{21})^2+(Z_{12}-Z_{22})^2]
\frac{1+\det^2Z}{2\det^2Z},\nn
\Delta^{Q,2}_2&=&2\frac{Z_{11}^2+Z_{21}^2}{\det^2Z}\ .
\eea
The magnetization dependence of $\Delta^{\rm Q,2}_{1,2}$ is shown
in Fig.~\ref{fig:su3exp}. 
\subsubsection{Dominant Power Law Correlations}
We are now in a position to identify the dominant power-law
correlations. In Fig.~\ref{fig:su3exp2} we plot the magnetization
dependence of the smallest exponents. We see that at low magnetizations
the quadrupolar correlation dominate, while for larger magnetizations
the transverse spin correlations are seen to decay slowest. The
cross-over between these two regimes occurs at $M\approx 0.258$.




\begin{thebibliography}{10}

\bibitem{spin_nematics1}
A.~F. Andreev and I.~A. Grishchuk, Sov. Phys. JETP {\bf 60},  267  (1984).

\bibitem{spin_nematics2}
A. Chubukov, J. Phys. Condens. Matter {\bf 2},  1593  (1990).

\bibitem{SatoruNakatsuji09092005}
S. Nakatsuji, Y. Nambu, H. Tonomura, O. Sakai, S. Jonas, C. Broholm, H.
  Tsunetsugu, Y. Qiu, and Y. Maeno, Science {\bf 309},  1697  (2005).

\bibitem{prl_triangularlattice}
A. L\"{a}uchli, F. Mila, and K. Penc, Phys. Rev. Lett. {\bf 97},  087205
  (2006).

\bibitem{tsunetsugu_arikawa}
H. Tsunetsugu and M. Arikawa, Journal of the Physical Society of Japan {\bf
  75},  083701  (2006).

\bibitem{springer_quantummagnetism}
U. Schollw\"{o}ck, J. Richter, D. Farnell, and R.~B. (Eds.), {\em Quantum
  Magnetism}, {\em Lecture Notes in Physics} (Springer-Verlag,
  Berlin/Heidelberg, 2004).

\bibitem{mila_zhang}
F. Mila and F.-C. Zhang, Eur. Phys. J. B {\bf 16},  7  (2000).

\bibitem{Cirac_PRL_spinhamiltonians_OptLatt}
J.~J. Garc\'\i{}a-Ripoll, M.~A. Martin-Delgado, and J.~I. Cirac, Phys. Rev.
  Lett. {\bf 93},  250405  (2004).

\bibitem{S.Trotzky01182008}
S. Trotzky, P. Cheinet, S. Folling, M. Feld, U. Schnorrberger, A.~M. Rey, A.
  Polkovnikov, E.~A. Demler, M.~D. Lukin, and I. Bloch, Science {\bf 319},  295
   (2008).

\bibitem{sun_hamiltonians}
A.~V. Gorshkov, M. Hermele, V. Gurarie, C. Xu, P.~S. Julienne, J. Ye, P.
  Zoller, E. Demler, M.~D. Lukin, and A.~M. Rey, Nature Physics {\bf 6},  289
  (2010).

\bibitem{uimin}
G.~V. Uimin, JETP Lett. {\bf 12},  225  (1970).

\bibitem{lai}
C.~K. Lai, J. Math. Phys. {\bf 15},  1675  (1974).

\bibitem{sutherland}
B. Sutherland, Phys. Rev. B {\bf 12},  3795  (1975).

\bibitem{takhtajan}
L. Takhtajan, Phys. Lett. A {\bf 87},  479  (1982).

\bibitem{babujian}
H. Babujian, Phys. Lett. A {\bf 90},  479  (1982).

\bibitem{PRL_AKLT}
I. Affleck, T. Kennedy, E.~H. Lieb, and H. Tasaki, Phys. Rev. Lett. {\bf 59},
  799  (1987).

\bibitem{barber1989}
M.~N. Barber and M.~T. Batchelor, Phys. Rev. B {\bf 40},  4621  (1989).

\bibitem{kluemper1989}
A. Kl\"{u}mper, Europhys. Lett. {\bf 9},  815  (1989).

\bibitem{Xian1993}
Y. Xian, Phys. Lett. A {\bf 183},  437  (1993).

\bibitem{fath1991}
G. F\'ath and J. S\'olyom, Phys. Rev. B {\bf 44},  11836  (1991).

\bibitem{fath1993}
G. F\'ath and J. S\'olyom, Phys. Rev. B {\bf 47},  872  (1993).

\bibitem{Schollwock1996}
U. Schollw\"ock, T. Jolic\oe{}ur, and T. Garel, Phys. Rev. B {\bf 53},  3304
  (1996).

\bibitem{buchta2005}
K. Buchta, G. F\'ath, O. Legeza, and J. S\'olyom, Phys. Rev. B {\bf 72},
  054433  (2005).

\bibitem{PRB_Lauchli}
A. L\"auchli, G. Schmid, and S. Trebst, Phys. Rev. B {\bf 74},  144426  (2006).

\bibitem{haldane1983}
F.~D.~M. Haldane, Phys. Lett. A {\bf 93},  464  (1983).

\bibitem{haldane1983PRL}
F.~D.~M. Haldane, Phys. Rev. Lett. {\bf 50},  1153  (1983).

\bibitem{fath_suto_2000}
G. F\'ath and A. S\"ut\ifmmode~\mbox{\H{o}}\else \H{o}\fi{}, Phys. Rev. B {\bf
  62},  3778  (2000).

\bibitem{parkinson1989}
J. Parkinson, J. Phys.: Condens. Matter {\bf 1},  6709  (1989).

\bibitem{okunishi1999rapid}
K. Okunishi, Y. Hieida, and Y. Akutsu, Phys. Rev. B {\bf 60},  6953(R)  (1999).

\bibitem{okunishi1999}
K. Okunishi, Y. Hieida, and Y. Akutsu, Phys. Rev. B {\bf 59},  6806  (1999).

\bibitem{fath_littlewood}
G. F\'ath and P.~B. Littlewood, Phys. Rev. B {\bf 58},  14709(R)  (1998).

\bibitem{white1992}
S.~R. White, Phys. Rev. Lett. {\bf 69},  2863  (1992).

\bibitem{schollwoeck2005}
U. Schollw\"ock, Rev. Mod. Phys. {\bf 77},  259  (2005).

\bibitem{kolezhuk_vekua2005}
A. Kolezhuk and T. Vekua, Phys. Rev. B {\bf 72},  094424  (2005).

\bibitem{harada_kawashima}
K. Harada and N. Kawashima, Phys. Rev. B {\bf 65},  052403  (2002).

\bibitem{prb_corboz}
P. Corboz, A.~M. L\"auchli, K. Totsuka, and H. Tsunetsugu, Phys. Rev. B {\bf
  76},  220404  (2007).

\bibitem{vekua2007}
T. Vekua, A. Honecker, H.-J. Mikeska, and F. Heidrich-Meisner, Phys. Rev. B
  {\bf 76},  174420  (2007).

\bibitem{Kecke2007}
L. Kecke, T. Momoi, and A. Furusaki, Phys. Rev. B {\bf 76},  060407  (2007).

\bibitem{Hikihara2008}
T. Hikihara, L. Kecke, T. Momoi, and A. Furusaki, Phys. Rev. B {\bf 78},
  144404  (2008).

\bibitem{PRB_Sudan}
J. Sudan, A. L\"uscher, and A.~M. L\"auchli, Phys. Rev. B {\bf 80},  140402
  (2009).

\bibitem{IanMcCulloch_VectorChiralOrder}
I.~P. McCulloch, R. Kube, M. Kurz, A. Kleine, U. Schollw\"ock, and A.~K.
  Kolezhuk, Phys. Rev. B {\bf 77},  094404  (2008).

\bibitem{book_CFT}
M. Henkel, {\em Conformal invariance and critical phenomena} (Springer, Berlin,
  1999).

\bibitem{calabrese_cardy}
P. Calabrese and J. Cardy, Journal of Statistical Mechanics: Theory and
  Experiment {\bf 2004},  P06002  (2004).

\bibitem{legeza_PRL}
\"O. Legeza, J. S\'olyom, L. Tincani, and R.~M. Noack, Phys. Rev. Lett. {\bf 99}, 087203 (2007). 

\bibitem{HeidrichMeisner2006}
F. Heidrich-Meisner, A. Honecker, and T. Vekua, Phys. Rev. B {\bf 74},  020403
  (2006).

\bibitem{Chubukov1991}
A.~V. Chubukov, Phys. Rev. B {\bf 44},  4693  (1991).

\bibitem{HeidrichMeisner2009}
F. Heidrich-Meisner, I.~P. McCulloch, and A.~K. Kolezhuk, Phys. Rev. B {\bf
  80},  144417  (2009).

\bibitem{schmidt2006}
K.~P. Schmidt, J. Dorier, A. L\"auchli, and F. Mila, Phys. Rev. B {\bf 74},
  174508  (2006).

\bibitem{fath2003}
G. F\'ath, Phys. Rev. B {\bf 68},  134445  (2003).

\bibitem{Konik2002}
R.~M. Konik and P. Fendley, Phys. Rev. B {\bf 66},  144416  (2002).

\bibitem{FZ}
A. Furusaki and S.-C. Zhang, Phys. Rev. B {\bf 60},  1175  (1999).

\bibitem{Lorenzo2002}
L. Campos~Venuti, E. Ercolessi, G. Morandi, P. Pieri, and M. Roncaglia, Int. J.
  Mod. Phys. B {\bf 16},  1363  (2002).

\bibitem{prb_friedrich}
A. Friedrich, A.~K. Kolezhuk, I.~P. McCulloch, and U. Schollw\"ock, Phys. Rev.
  B {\bf 75},  094414  (2007).

\bibitem{karlo_communication}
K. Penc, private communication.

\bibitem{tsvelik_book}
A.~M. Tsvelik, {\em Quantum Field Theory in Condensed Matter Physics}
  ({Cambridge University Press}, Cambridge, 2003).

\bibitem{Ejima}
S. Ejima, M.~J. Bhaseen, M. Hohenadler, F.~H.~L. Essler, H. Fehske, and B.~D.
  Simons, Phys. Rev. Lett. {\bf 106}, 015303 (2011).  

\bibitem{lecheminant_prl}
P. Lecheminant, E. Boulat, and P. Azaria, Phys. Rev. Lett. {\bf 95},  240402
  (2005).

\bibitem{wu_prl}
C. Wu, Phys. Rev. Lett. {\bf 95},  266404  (2005).

\bibitem{lecheminant_nuclphysb}
P. Lecheminat, E. Boulat, and P. Azaria, Nucl. Phys. B {\bf 798},  443  (2008).

\bibitem{capponi_rapid}
S. Capponi, G. Roux, P. Azaria, E. Boulat, and P. Lecheminant, Phys. Rev. B
  {\bf 75},  100503(R)  (2007).

\bibitem{roux2009}
G. Roux, S. Capponi, P. Lecheminant, and P. Azaria, Eur. Phys. J. B {\bf 68},
  293  (2009).

\bibitem{bonnes2011}
L. Bonnes and S. Wessel, Phys. Rev. Lett. {\bf 106}, 185302 (2011).

\bibitem{amt}
A.~M. Tsvelik, Phys. Rev. B {\bf 42},  10499  (1990).

\bibitem{SGM1}
F. Haldane, J. Phys. A {\bf 15},  507  (1982).

\bibitem{SGM2}
J.-S. Caux and A. Tsvelik, Nucl. Phys. B {\bf 474},  715  (1996).

\bibitem{Sitte}
M. Sitte, A. Rosch, J.~S. Meyer, K.~A. Matveev, and M. Garst, Phys. Rev. Lett. {\bf 102}, 176404 (2009). 
 
\bibitem{shura}
A.~A. Nersesyan and A.~M. Tsvelik, Phys. Rev. Lett. {\bf 78},  3939  (1997).

\bibitem{Affleck1991}
I. Affleck, Phys. Rev. B {\bf 43},  3215  (1991).

\bibitem{babujian_nuclphysB}
H. Babujian, Nucl. Phys. B {\bf 215},  317  (1983).

\bibitem{TsvelikNucPhysB}
A. Tsvelik, Nucl. Phys. B {\bf 305},  675  (1988).

\bibitem{strings1}
F. Alcaraz and M. Martins, J. Phys. A {\bf 22},  1829  (1989).

\bibitem{strings2}
F. Alcaraz and M. Martins, J. Phys. A {\bf 23},  1439  (1990).

\bibitem{strings3}
H. Frahm and N.-C. Yu, J. Phys. A {\bf 23},  2115  (1990).

\bibitem{strings4}
H. Frahm, N.-C. Yu, and M. Fowler, Nucl. Phys. B {\bf 336},  396  (1990).

\bibitem{vladbook}
V. Korepin, A. Izergin, and N. Bogoliubov, {\em Quantum Inverse Scattering
  Method, Correlation Functions and Algebraic Bethe Ansatz} (Cambridge
  University Press, Cambridge, 1993).

\bibitem{Cardy86}
J.~L. Cardy, Nucl. Phys. B {\bf 270},  186  (1986).

\bibitem{FSspectrum1}
F. Woynarovich, J. Phys. A {\bf 22},  4243  (1989).

\bibitem{FSspectrum2}
H. Frahm and V.~E. Korepin, Phys. Rev. B {\bf 42},  10553  (1990).

\bibitem{book_hubbardmodel}
F.~H.~L. Essler, H. Frahm, F. G\"ohmann, A. Kl\"umper, and V.~E. Korepin, {\em
  The One-Dimensional Hubbard Model} (Cambridge University Press, Cambridge,
  2005).

\bibitem{FrKo91}
H. Frahm and V.~E. Korepin, Phys. Rev. B {\bf 43},  5653  (1991).

\bibitem{EF99}
F.~H.~L. Essler and H. Frahm, Phys. Rev. B {\bf 60},  8540  (1999).

\end{thebibliography}

\end{document}